\documentclass[structabstract]{aa}  
\usepackage{graphicx}
\usepackage{txfonts}
\newcommand{\new}[1]{{#1}}

\begin{document}
   \title{Massive open star clusters using the VVV survey\thanks{Based on observations with ISAAC, VLT, ESO (programme 087.D-0341A), New Technology Telescope at ESO's La Silla Observatory (programme 087.D-0490A) and with the Clay telescope at the Las Campanas Observatory (programme CN2011A-086). Also based on data from the VVV survey (programme 172.B-2002).}}

   \subtitle{II. Discovery of six clusters with Wolf-Rayet stars}

   \author{A.-N. Chen\'e\inst{1,2}, J. Borissova\inst{1,3}, C. Bonatto\inst{4}, D.~J. Majaess\inst{5}, G. Baume\inst{6}, J.~R.~A. Clarke\inst{1,3,7}, R. Kurtev\inst{1}, O. Schnurr\inst{8}, J.-C. Bouret\inst{9}, M. Catelan\inst{3,10}, J.~P. Emerson\inst{11}, C. Feinstein\inst{6}, D. Geisler\inst{2}, R. de Grijs\inst{12}, A. Herv\'e \inst{13}, V.~D. Ivanov\inst{14}, M.~S.~N. Kumar\inst{15},  P. Lucas\inst{7}, L. Mahy\inst{13}, F. Martins\inst{16}, F. Mauro\inst{2}, D. Minniti\inst{3,10,17,18} and C. Moni Bidin\inst{2,3}} \authorrunning{Chen\'e et al.}
   \titlerunning{Discovery of six clusters with Wolf-Rayet stars in VVV}
                  %1
   \institute{Departamento de F\'{\i}sica y Astronom\'{\i}a, Universidad de Valpara\'{\i}so, Av. Gran Breta\~na 1111, Playa Ancha, Casilla 5030, Chile\\
              \email{andrenicolas.chene@gmail.com}
         \and %2
             Departamento de Astronom\'{\i}a, Universidad de Concepci\'on, Casilla 160-C, Chile
         \and %3
             The Milky Way Millennium Nucleus, Av. Vicu\~{n}a Mackenna 4860, 782-0436 Macul, Santiago, Chile
         \and %4
             Universidade Federal do Rio Grande do Sul, Departamento de Astronomia CP 15051, RS, Porto Alegre, 91501-970, Brazil
         \and %5
            Saint Mary's University, Halifax, Nova Scotia, Canada
         \and %6
             Facultad de Ciencias Astron\'omicas y Geof\'{\i}sicas (UNLP), Instituto de Astrof\'{\i}sica de La Plata (CONICET, UNLP), Paseo del Bosque s/n, La Plata, Argentina
         \and %7
             University of Hertfordshire, Hatfield, AL10 9AB, UK
         \and %8
             Leibniz-Institut f\"ur Astrophysik Potsdam (AIP), An der Sternwarte 16, 14482, Potsdam, Germany
         \and %9
             Laboratoire d'Astrophysique de Marseille -- LAM, Universit\'e d'Aix-Marseille \& CNRS, UMR7326, 38 rue F. Joliot-Curie, 13388 Marseille Cedex 13, France
         \and %10
             Pontificia Universidad Cat\'olica de Chile, Facultad de F\'{\i}sica, Departamento de Astronom\'{\i}a y Astrof\'{\i}sica, Av. Vicu\~{n}a Mackenna 4860, 782-0436 Macul, Santiago, Chile
         \and %11
            11 School of Physics and Astronomy, Queen Mary University of London, London E1 4NS, United Kingdom
         \and %12
             Kavli Institute for Astronomy and Astrophysics, Peking University, Yi He Yuan Lu 5, Hai Dian District, Beijing 100871, China
         \and %13
             Groupe d'Astrophysique des Hautes Energies, Institut d'Astrophysique et de G\'eophysique, Universit\'e de Li\`ege, All\'ee du 6 Ao\^ut, Batiment B5c, 4000 Li\`ege, Belgium 
         \and %14
             European Southern Observatory, Ave. Alonso de Cordova 3107, Casilla 19, Chile
         \and %15
             Centro de Astrofisica da Universidade do Porto, Rua das Estrelas, 4150-762 Porto, Portugal
         \and %16
             LUPM-UMR5299, CNRS \& Universit\'e Montpellier II, Place Eug\`ene Bataillon, F-34095, Montpellier Cedex 05, France 
         \and %17
             Vatican Observatory, V00120 Vatican City State
         \and %18
             Department of Astrophysical Sciences, Princeton University, Princeton NJ 08544-1001
%             \email{c.ptolemy@hipparch.uheaven.space}
%             \thanks{The university of heaven temporarily does not
%                     accept e-mails}
             }

   \date{Received July 26, 2012; accepted XXXX XX, 2012}

% \abstract{}{}{}{}{} 
 \abstract
% context 
{The ESO Public Survey ``VISTA Variables in the V\'ia L\'actea'' (VVV) provides deep multi-epoch infrared observations for an unprecedented 562 sq. degrees of the Galactic bulge, and adjacent regions of the disk. In this survey nearly 150 new open clusters and cluster candidates have been discovered.}
% aims 
{This is the second in a series of papers about young, massive open clusters observed using the VVV survey. We present the first study of six recently discovered clusters. These clusters contain at least one newly discovered Wolf-Rayet (WR) star.}
% methods 
{Following the methodology presented in the first paper of the series, wide-field, deep $JHK_{\rm s}$ VVV observations, combined with new infrared spectroscopy, are employed to constrain fundamental parameters for a subset of clusters.}
% results 
{We affirm that the six studied stellar groups are real young (2--7\,Myr) and massive (between 0.8 and 2.2 $10^{3}$\,$M_\odot$) clusters. They are highly obscured ($A_V\sim 5-24$\,mag) and compact (1--2\,pc). In addition to WR stars, two of the six clusters also contain at least one red supergiant star. We claim the discovery of 8 new WR stars, and 3 stars showing WR-like emission lines which could be classified WR or OIf. Preliminary analysis provides initial masses of $\sim$30--50\,$M_\odot$ for the WR stars. Finally, 
we discuss the spiral structure of the Galaxy using as tracers the six new clusters together with the previously studied VVV clusters.}
% conclusions 
{}


   \maketitle
%
%________________________________________________________________

\section{Introduction}

Borissova et al. (\cite{Bo11}) discovered 96 new open clusters and stellar groups in the Galactic disk area that was covered by the infrared (IR) VISTA Variables in the V\'{\i}a L\'actea (VVV) survey (Minniti et al. \cite{Mi10}, Saito et al. \cite{Sa12}), one of six Europeam Southern Observatory (ESO) Public Surveys carried out with the new 4 m Visible and Infrared Survey Telescope for Astronomy (VISTA). Most of the new cluster candidates are faint and compact (characterized by small angular sizes), highly reddened, and younger than 5\,Myr, thus demonstrating the potential of the VVV survey to find new open clusters, especially those hiding in dusty regions.

In the year following this discovery, our group has initiated a near-IR spectroscopic follow-up campaign to confirm their cluster nature, and to derive the spectral types and distances of the brighter cluster members. A total of 27 new VVV cluster candidates have thus far been observed, as well as $\sim$40 other known cluster candidates present within the VVV area. Our final objective is the characterization of a large sample of star clusters using homogeneous data and analysis methods. Our results for four known clusters, Danks\,1, Danks\,2, RCW\,79 and DBS\,132, were presented by Chen\'e et al. (\cite{Ch12}, hereafter Paper\,I). Here, we use the same method as in Paper\,I, but applied to six of the new VVV cluster candidates that have not been studied before and in which spectroscopic observations have revealed the presence of Wolf-Rayet (WR) stars. Studies of the other clusters will be presented in a series of papers in preparation.

The study of young massive clusters containing WR stars is a key to understand the evolution of massive stars. Typically, WR stars have masses of 10--25 $M_\odot$ and they are believed to have descended from O-type stars. They exhibit strong, broad emission lines of He and N (WN spectral type) or He, C and O (WC stars). They spend $\sim$10\% of their $\sim$5\,Myr overall lifetime as WR stars (Meynet \& Maeder \cite{Me05}). At solar metallicity and in the absence of rotation, O stars with an initial mass of at least $\sim$25\,$M_\odot$ can evolve into WR stars, but most have initial masses of 40--50\,$M_\odot$, allowing them to reach the WC sequence (Crowther \cite{Cr07}). Consequently, if we assume that the mass of the most massive star formed in a cluster is constrained by the mass of the cluster itself (as proposed by Weidner et al. \cite{We10}), initial total cluster masses must be $\sim$500\,$M_\odot$, or even $\sim$1000\,$M_\odot$ if they contain at least one WN or WC star, respectively (based on the analytical model of Weidner \& Kroupa \cite{We04}). On the other hand, if there is no real relation between the mass of the most massive member and the total mass of a cluster (Eldridge \cite{El12}), lower initial total cluster masses can be accommodated.

If main-sequence (MS) and evolved O stars are detected together with WR stars, tentative evolutionary sequences can be defined. If the overall population was formed in a burst of star formation, WR stars have progenitorsÊ with spectral \new{types} earlier than the O stars still present in the cluster. Consequently, the nature of WR progenitors can be constrained (e.g. Martins et al. \cite{Ma08} in the Arches cluster) and direct connections between MS O stars and evolved WR stars are established. In addition, the identification of MS O stars is a crucial piece of information: it shows the position of the turn-off of the entire population, providing constraints on the age (e.g. Martins et al. \cite{Ma07}).
%[+ d'autres, sur les amas globulaires par ex, voir Iben]
This provides upper limits on the main sequence lifetime of the progenitors of WR stars. If red supergiants are present at the same time as WR stars, as is the case in only two clusters (the central cluster and Westerlund 1) this strongly constrains the link between hot and cool evolved massive stars in the upper HR diagram, potentially revealing evolutionary sequences between certain WR stars and red supergiants. Obtaining information of clusters of different ages is necessary to establish the evolutionary sequences followed by stars of different masses. Very young clusters have only very massive stars off the main sequence, while older (i.e. 3 to 10\,Myr) contain evolved O stars of lower and lower masses as the age increases. In short, young massive clusters are a gold mine to understand stellar evolution.

 Finally, once the distances to these clusters have been obtained, they can be used to analyse the Galaxy's spiral structure. Indeed, the unprecedented deepness and resolution of the data from the VVV survey allow us to extend the current sample of known open clusters towards the Milky WayÕs highly obscured region. The study of the VVV clusters hence opens a window towards unexplored parts of the Galaxy.

%.  Also, they make great targets to look for ejection history, which gives precious information about previous stages. 
%\begin{landscape}
\begin{figure*}[ht]
  \centering
   \includegraphics[width=7.cm]{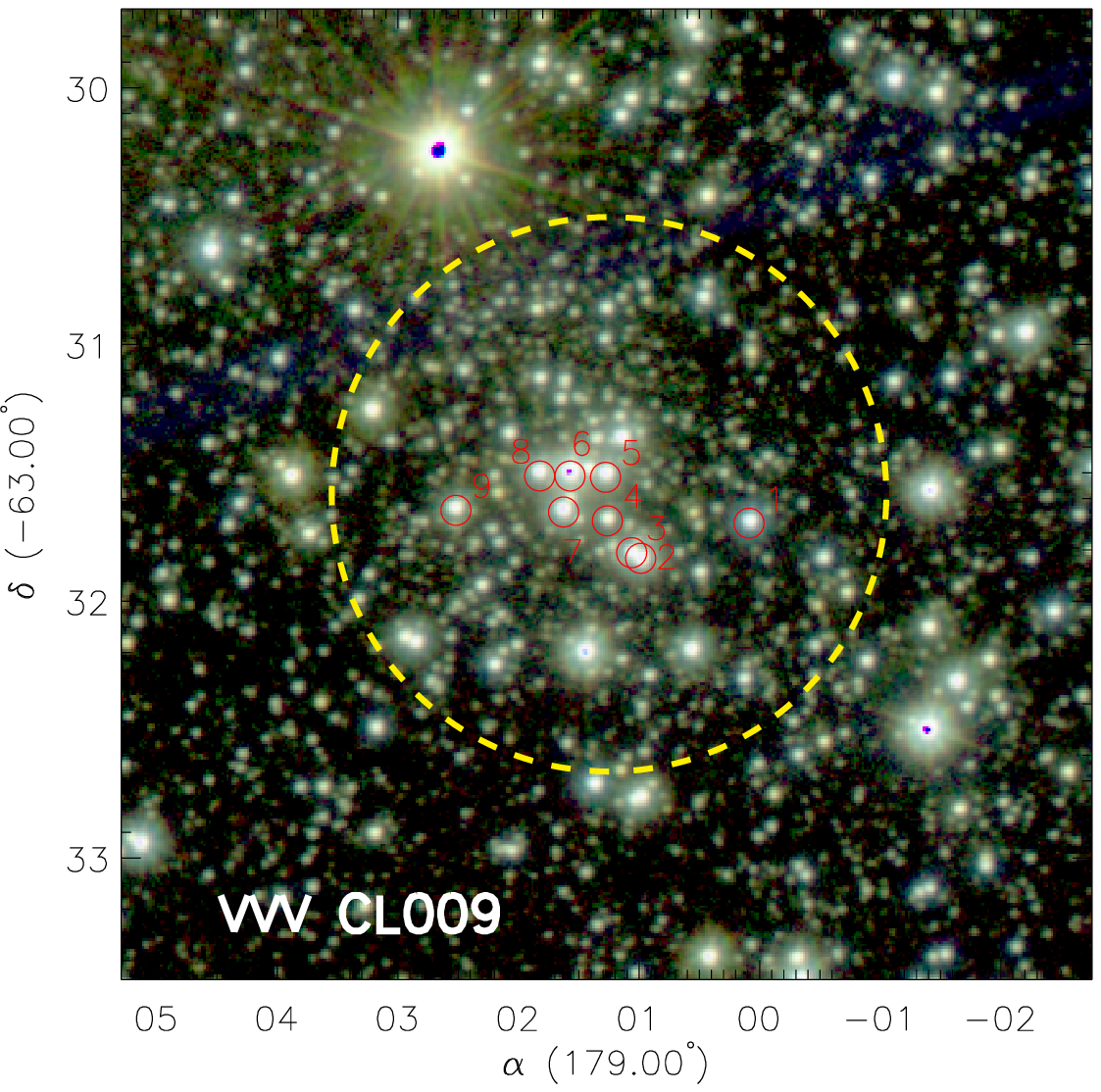}
   \includegraphics[width=7.cm]{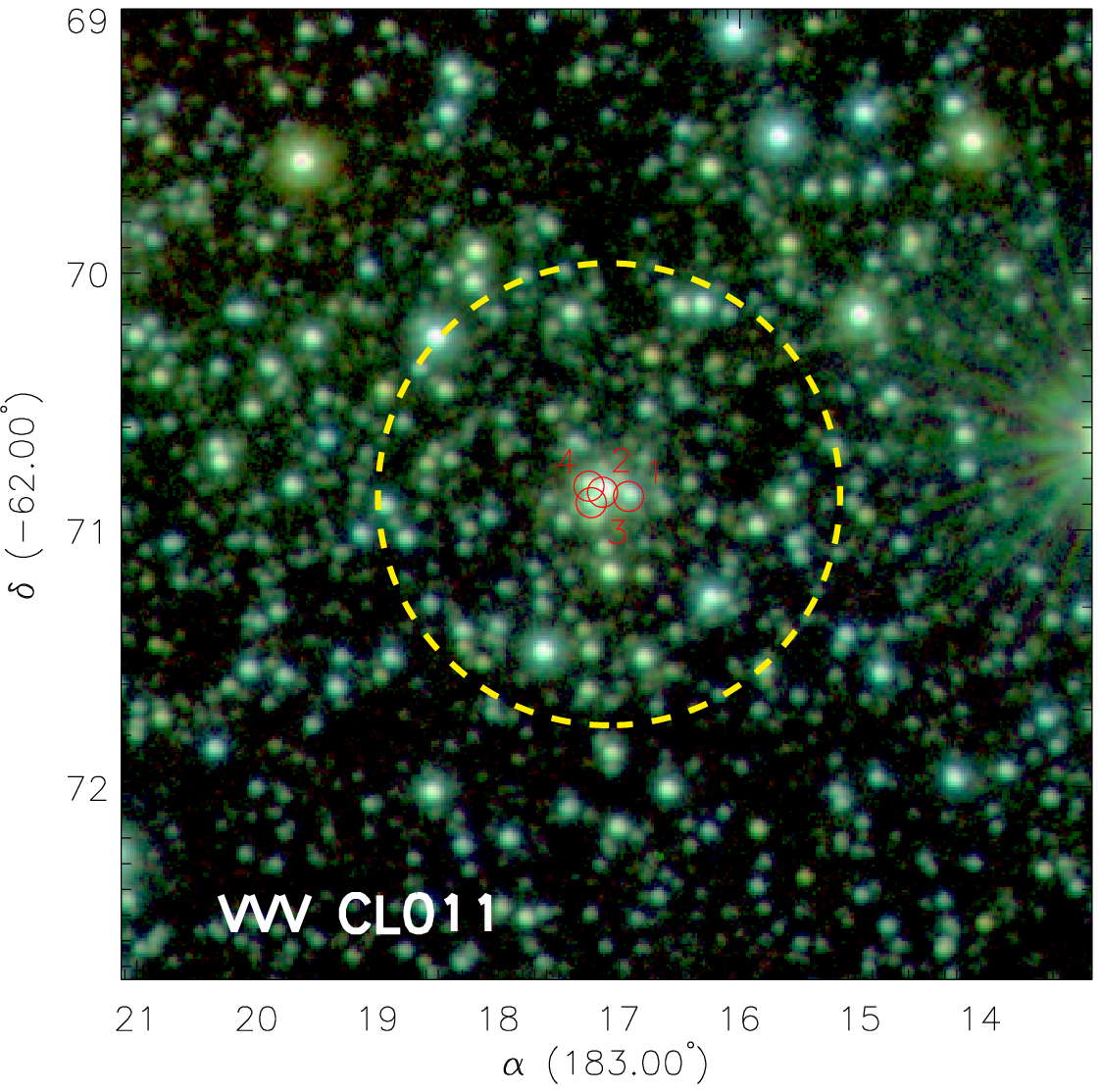}
   \includegraphics[width=7.cm]{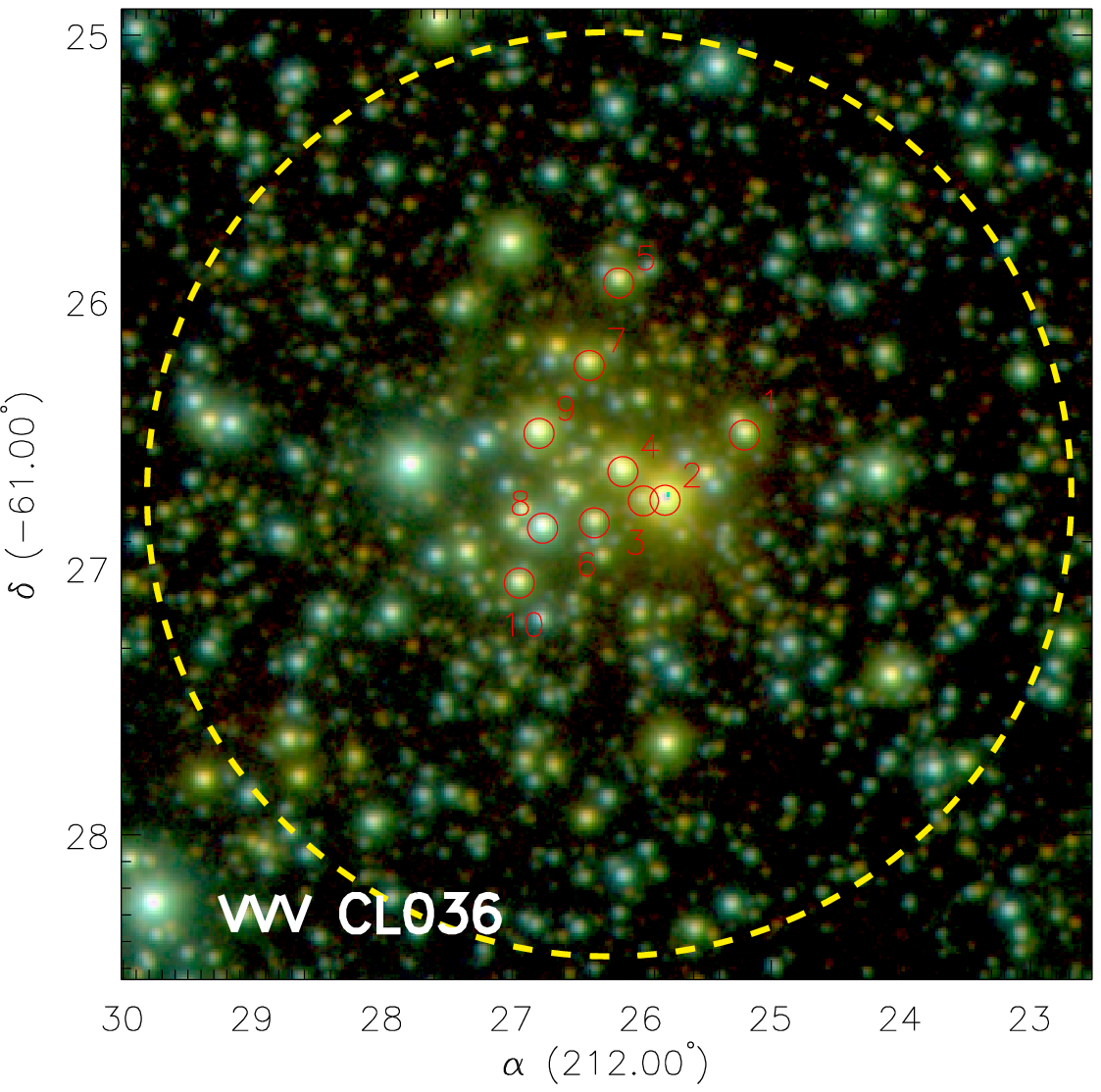}
   \includegraphics[width=7.cm]{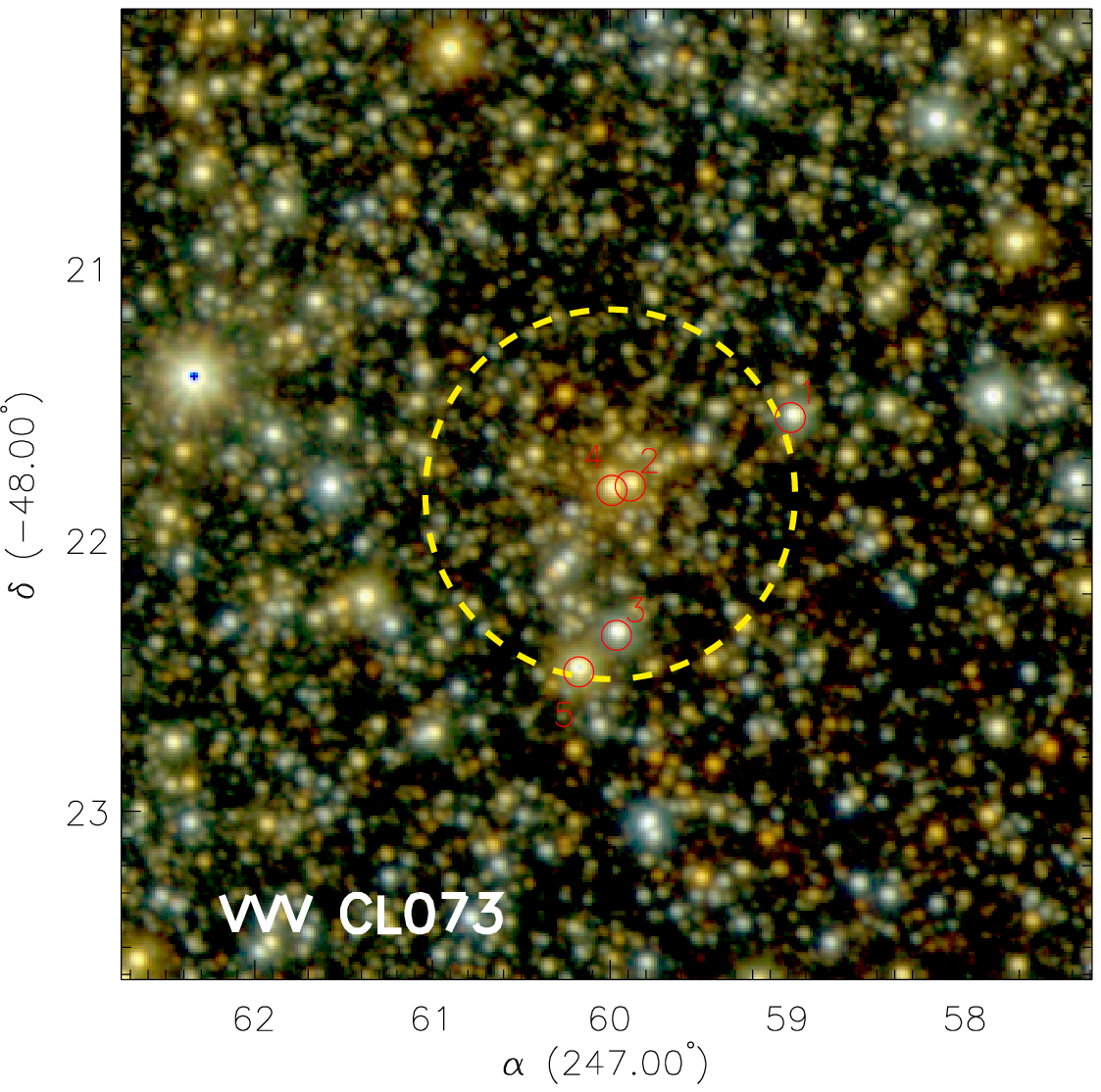}
   \includegraphics[width=7.cm]{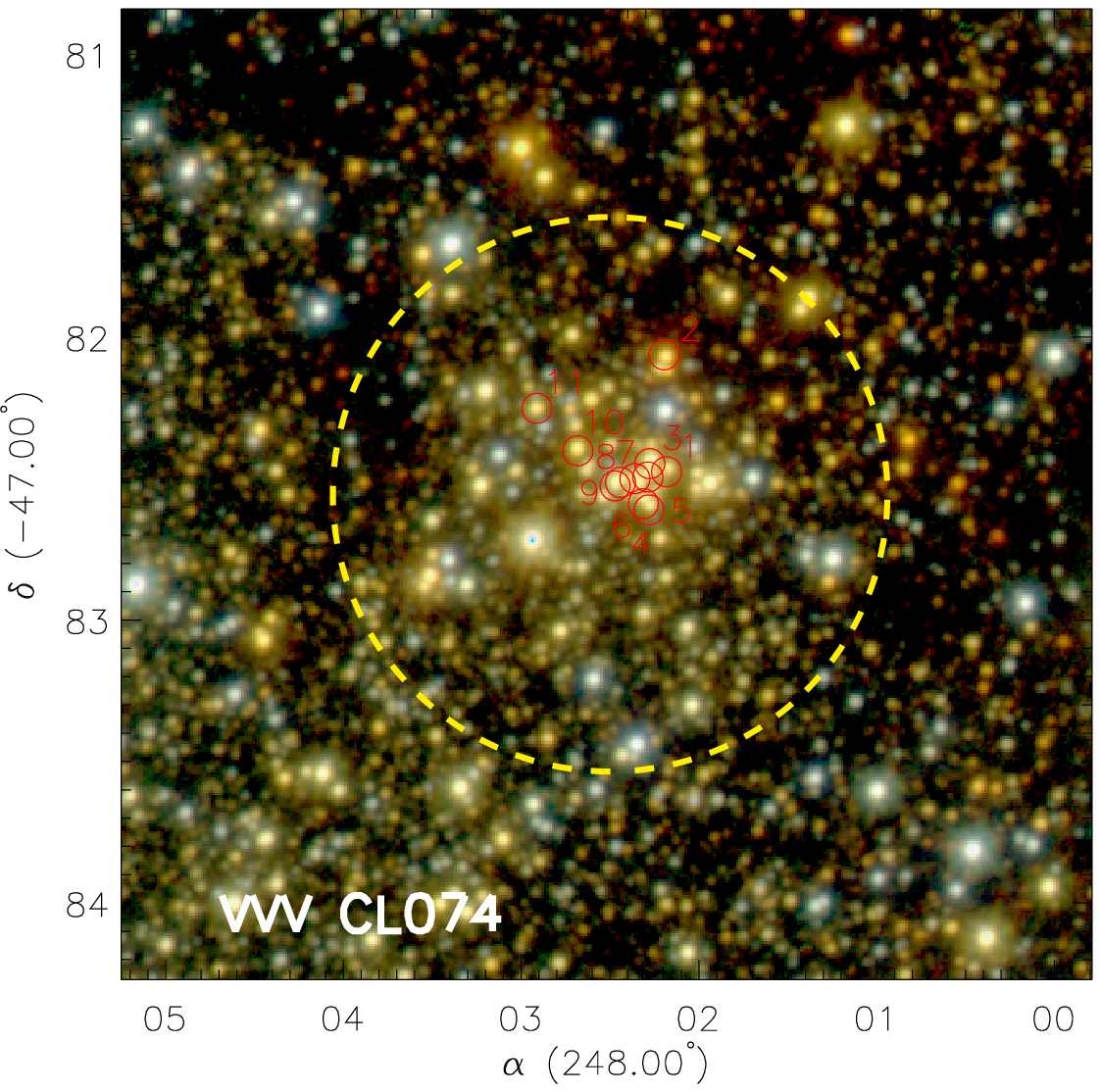}
   \includegraphics[width=7.cm]{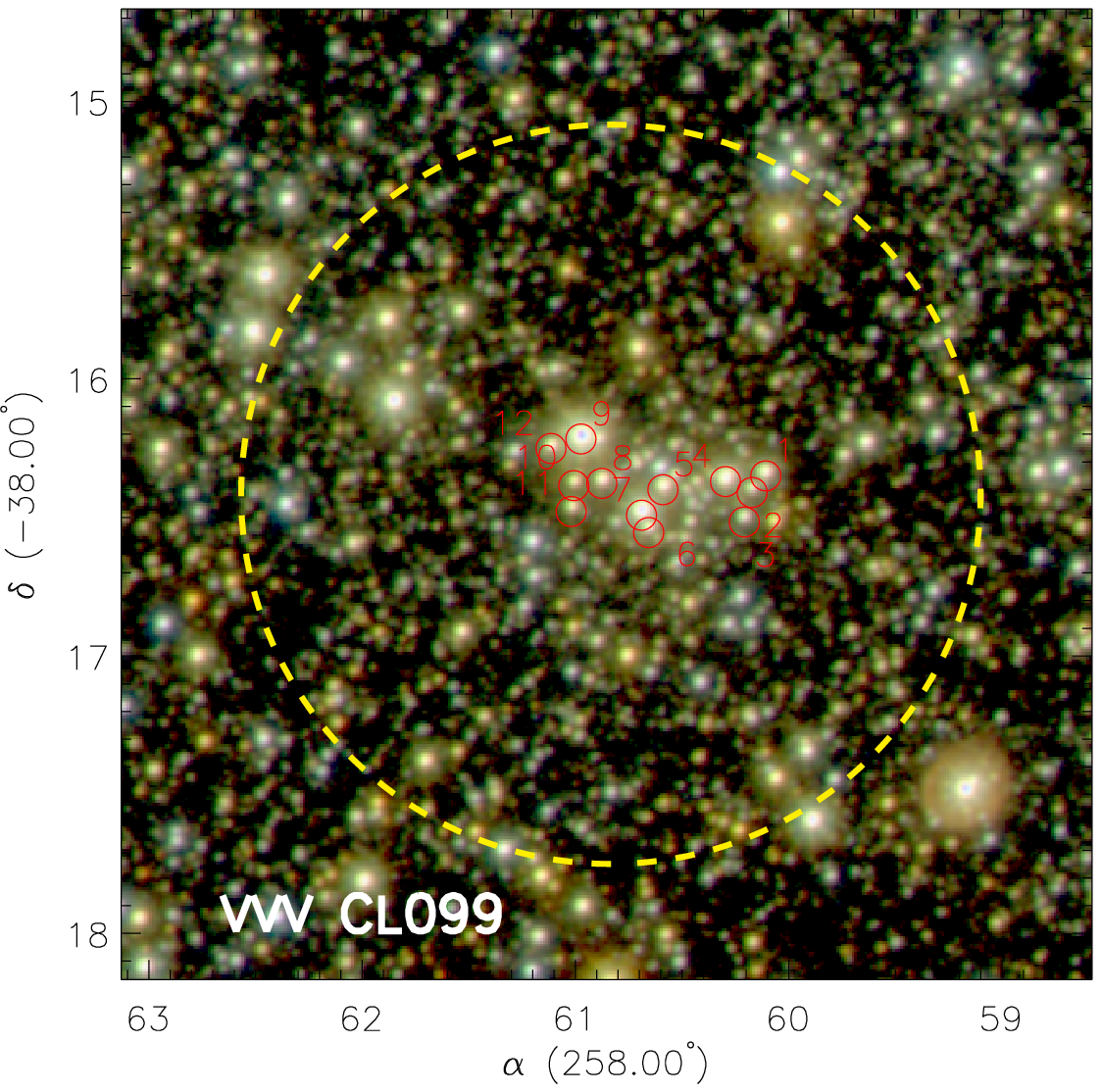}
   \caption{$JHK_{\rm s}$ \new{false}-colour images of the six clusters. Stars marked with red circles were observed using near-IR spectrographs. Yellow dashed circles indicate the angular sizes of the clusters (see Section\,\ref{AngSiz}). Coordinates are given in the J2000 system.}
              \label{FOV}%
\end{figure*}
%\end{landscape}

We present the observational data in Section\,\ref{obs}, summarize our method in Section\,\ref{Met} and present the details of the clusters in Section\,\ref{Results}. We then discuss the likelihood that the six stellar groups are real clusters, as well as the Galaxy's spiral structure as traced by the open clusters observed in the VVV data and the properties of 11 newly discovered WR stars in Section\,\ref{Discussion} before we summarize this work in Section\,\ref{Summary}.

\section{Observations}\label{obs}

\subsection{Near-IR photometry}\label{obs_phot}

\subsubsection{VVV data and photometry}

We downloaded the stacked images of the individual 2048$\times$2048 pixel exposures containing the six clusters discussed in this paper from the VISTA Science Archive (VSA) website.\footnote{http://horus.roe.ac.uk/vsa/} The names and coordinates of our sample clusters, as well as the relevant VVV fields (see Saito et al. \cite{Sa12} for more details about VVV data), are listed in Table\,\ref{TabCl}.

\begin{table*}
\caption{Sample clusters.}
\label{TabCl}      % is used to refer this table in the text
\centering                          % used for centering table
\begin{tabular}{ccccrccc}
\hline\hline
\noalign{\smallskip}
VVV&      R.A.    &   Dec    &   $l$   & \multicolumn{1}{c}{$b$}    &    size    & frames & VVV field\\ 
CL  & (J2000) & (J2000) & (deg) & \multicolumn{1}{c}{(deg)} & (arcmin)& & \\ 
\noalign{\smallskip}
\hline
\noalign{\smallskip}
009 & 11:56:03.18 & --63:18:42.19 & 296.7605 & --1.1055 & $15.5 \times 11.6$ & $2 \times J$, $2 \times H$, $12 \times K_{\rm s}$ & d002\\
011 & 12:12:41.22 & --62:42:33.52 & 298.5069 & --0.1710 & $15.5 \times 11.6$ & $2 \times J$, $2 \times H$, $12 \times K_{\rm s}$ &d041\\
036 & 14:09:03.95 & --61:15:58.14 & 312.1238 &   0.2125 & $23.2 \times 11.6$ & $3 \times J$, $3 \times H$, $18 \times K_{\rm s}$ & d088\\
073 & 16:30:24.20 & --48:13:00.44 & 335.8939 &   0.1326 & $23.2 \times 15.5$ & $4 \times J$, $4 \times H$, $40 \times K_{\rm s}$ & d105\\
074 & 16:32:05.86 & --47:49:30.50 & 336.3733 &   0.1944 & $15.5 \times 11.6$ & $2 \times J$, $2 \times H$, $22 \times K_{\rm s}$ & d105\\
099 & 17:14:25.78 & --38:09:58.28 & 348.7266 &   0.3233 & $15.5 \times 11.6$ & $2 \times J$, $2 \times H$, \,\,\,$6 \times K_{\rm s}$& d114\\
\noalign{\smallskip}
\hline
\end{tabular}
\tablefoot{Columns include the VVV\,CL name, coordinates (R.A. and Dec; J2000), the fields of view of the regions of interest, the number of frames in the different filters, and the name of the VVV field where the clusters are located.}
\end{table*}

These fields were acquired in the $JHK_{\rm s}$ filters in March and April 2010 with the VIRCAM camera mounted on the VISTA 4 m telescope at Paranal Observatory (Emerson \& Sutherland \cite{Em10}) and reduced at CASU\footnote{http://casu.ast.cam.ac.uk/} using the VIRCAM pipeline v1.1 (Irwin et al. \cite{Ir04}). The total exposure time of each of the images was 40\,s, with two images per filter on average. For a detailed description of the observing strategy, see Minniti et al. (\cite{Mi10}). During the observations the weather conditions met all survey constraints for seeing ($<0.8''$), airmass ($<1.5$), sky transparency (clear) and Moon distance (Minniti et al. \cite{Mi10}), and the quality of the data was satisfactory (i.e., not affected by any artificial features related to problems caused by telescope or detector issues). $JHK_{\rm s}$ \new{false}-colour images of the six clusters are shown in Figure~\ref{FOV}. Additional 8\,s $K_{\rm s}$-band images were also obtained to discover and monitor variable stars in these fields. Unfortunately, to date fewer than 10 epochs have been observed for our fields of interest, which is insufficient for a thorough investigation of the presence and properties of binaries and pulsating stars. 

Similarly to Paper\,I, stellar photometry was performed by employing the VVV-SkZ\_pipeline's (Mauro et al. \cite{Ma12}) automated software based on ALLFRAME (Stetson \cite{St94}), optimized for VISTA point-spread-function photometry. Extensive details about the many steps required to obtain photometric measurements are described in section\,2.2 of Moni-Bidin et al. (\cite{Mo11}). The photometric accuracy reaches $\sigma=$\,0.05\,mag for the brightest ($J=9.0$\,mag and/or $K_{\rm s}=9.0$\,mag) and the faintest ($J=20.0$\,mag and/or $K_{\rm s}=18.0$\,mag) stars, and reaches $\sigma=$\,0.01\,mag for stars with intermediate brightnesses. \new{2MASS photometry was used for absolute flux calibration in the $J$, $H$ and $K_{\rm s}$ bands using stars with $12.5<J<14.5$~mag, $11.5<H<13$~mag and $11<K_{\rm s}<12.5$~mag.} We opted to work in the 2MASS system rather than sticking to the natural VISTA system, since most of the models we use in this study, as well as most other studies of massive clusters, employ the 2MASS photometric system. For stars brighter than $K_{\rm s}$=9.0~mag we simply adopted the 2MASS magnitudes.

\subsubsection{Statistical decontamination}

Statistical field-star decontamination was performed as described in Paper\,I, using the algorithm of Bonatto \& Bica (\cite{Bo10}). This algorithm divides the full range of magnitudes and colours into a grid of cells with axes along $K_{\rm s}$, ($H-K_{\rm s}$) and ($J-K_{\rm s}$). For each cell, it estimates the expected number density of member stars in the cluster by subtracting the respective field-star number density. As a photometric quality constraint, the algorithm rejects stars with uncertainties in $K_{\rm s}$ and colours in excess of 0.2~mag. The comparison fields we used for the decontamination depended on the projected distribution of individual stars, clusters or clumpy extinction due to dark clouds in an image. Examples include a ring around the cluster or some other region selected in its immediate vicinity. To avoid biases introduced by the subjective choice of comparison field, we compared many different geometries, positions and sizes of fields. Since the VVV catalogue is well populated, the differences in the final colour--magnitude diagram (CMD) were marginal.

\subsubsection{Completeness}

To calculate the completeness of our catalogue, we created luminosity functions for the clusters in bins of 0.5 $K_{\rm s}$-band magnitudes. For each magnitude bin, we added artificial stars to the cluster image within certain magnitude ranges. The relevant completeness fraction was calculated by recording the recovered fraction of artificial stars per unit input magnitude. Stellar detection was done using DAOphot, following a similar approach as in the VVV-SkZ\_pipeline.

\begin{figure*}[ht]
  \centering
   \includegraphics[width=6.0cm]{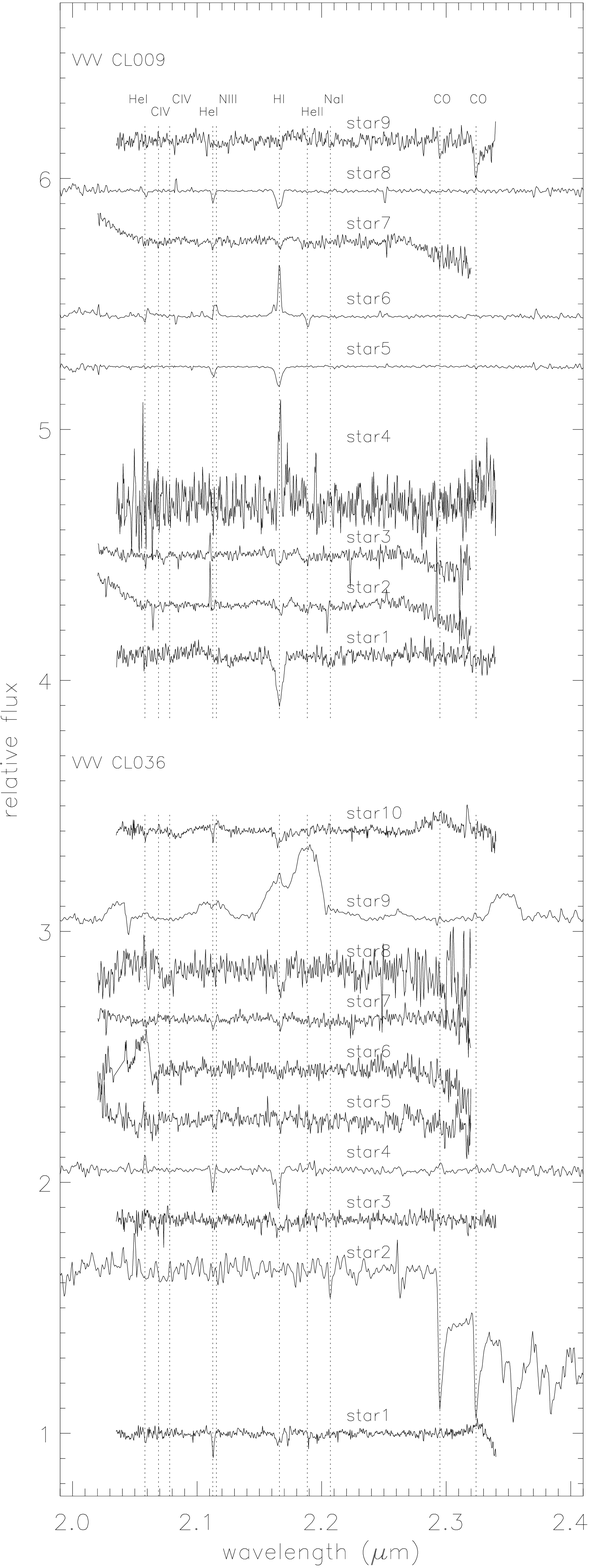}
   \includegraphics[width=	6.0cm]{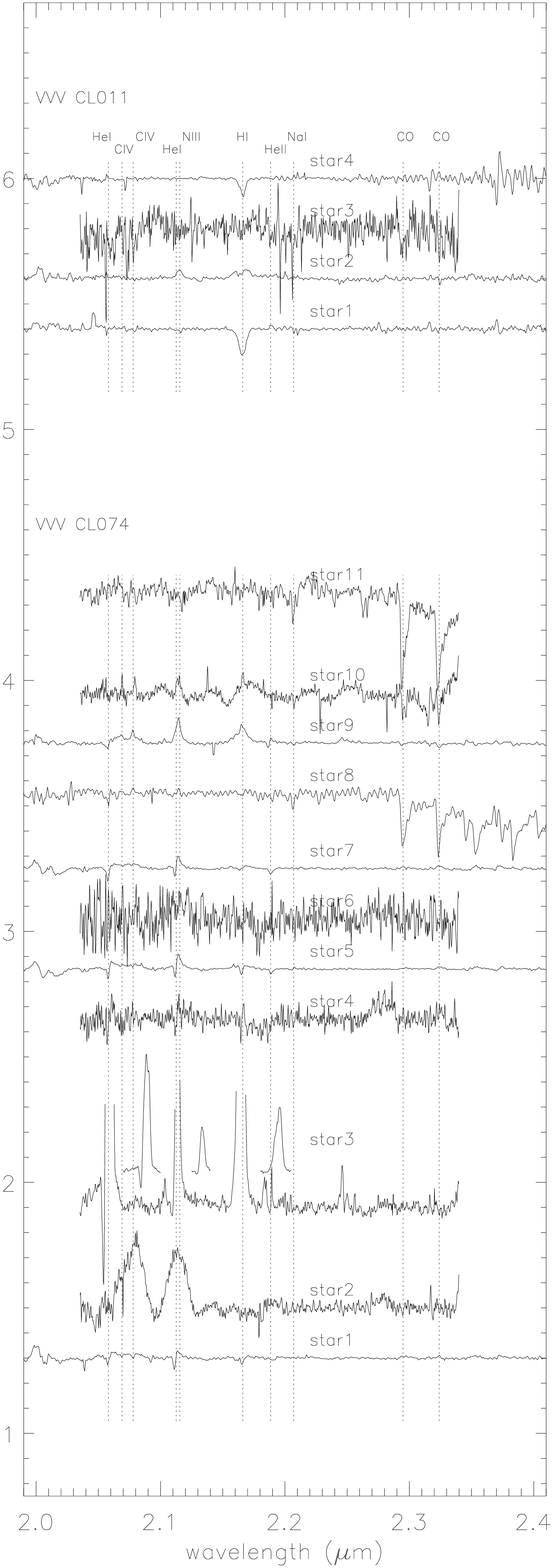}
   \includegraphics[width=	6.0cm]{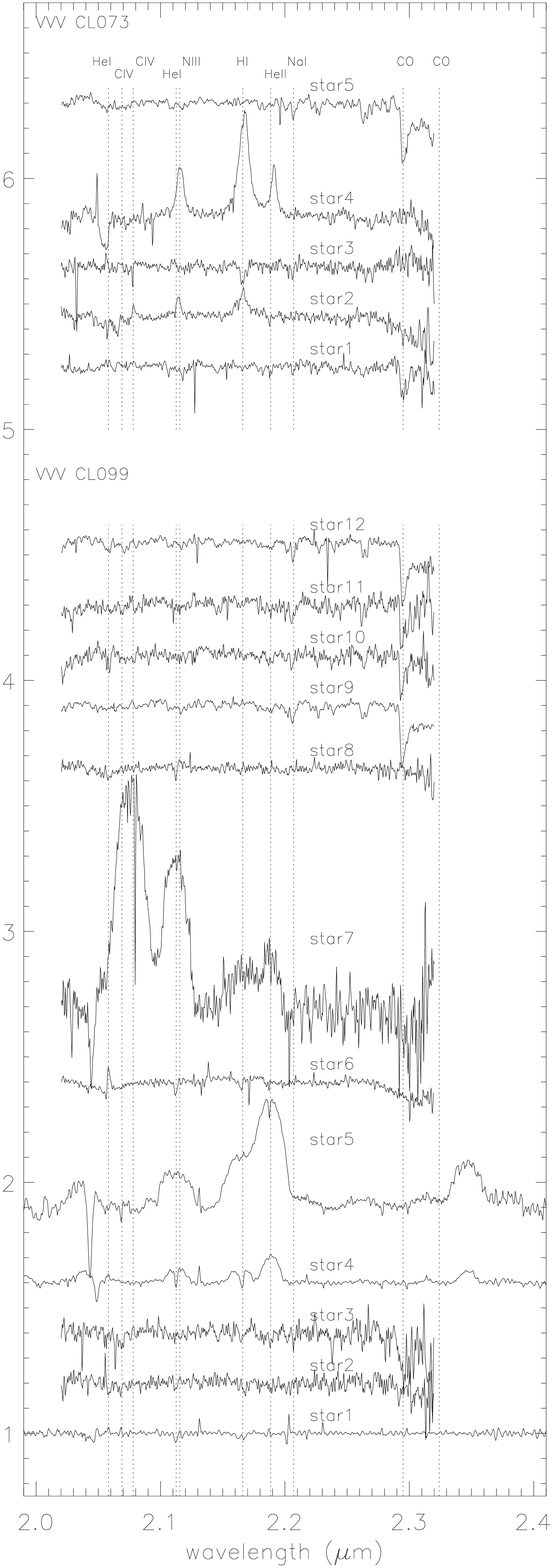}
   \caption{Stellar spectra in the area of the six clusters. Hydrogen and helium recombination lines, as well as C{\sc iv}, Na{\sc i} and CO lines are marked with vertical dotted lines. The wavelength range displayed depends on the spectrograph used.}
              \label{FigSp1}%
\end{figure*}

\subsection{Near-IR spectroscopy}

\subsubsection{Observations}

We collected spectra of most of the brightest members in the six clusters using the IR spectrographs and imaging cameras SofI in long-slit mode on the New Technology Telescope (NTT) at ESO's La Silla Observatory, OSIRIS at the Southern Observatory for Astrophysical Research (SOAR) and ISAAC at the Very Large Telescope at Paranal (ESO), all based in Chile. These stars are marked with red circles in Figure~\ref{FOV}. The instrument set-ups are summarized in Table\,\ref{TabSetUp}. Total exposure times were typically 200--400\,s for the brightest stars and 1200\,s for the faintest.

\begin{table*}
\caption{Near-IR spectrograph log of observations and set-ups.}
\label{TabSetUp}      % is used to refer this table in the text
\centering                          % used for centering table
\begin{tabular}{lllcccc}
\hline\hline
\noalign{\smallskip}
 \multicolumn{1}{c}{Telescope} &  \multicolumn{1}{c}{Instrument} &  \multicolumn{1}{c}{Dates} & slit size & Range & Resolution & Pixel scale\\ 
 \multicolumn{1}{c}{} &  \multicolumn{1}{c}{} &  \multicolumn{1}{c}{} & ($''$) & ($\mu$m) & $R$ & \AA\ pix$^{-1}$ \\ 
\noalign{\smallskip}
\hline
\noalign{\smallskip}
VLT 8.2m & ISAAC & 22--23 Apr 2011 & 0.3 & 1.98--2.43 & 3000 & 7.09\\
NTT 3.6m & SofI & 02--03 May 2012 & 0.6 & 2.02--2.32 & 2200 & 4.66 \\
SOAR 4.1m & OSIRIS & 06--07 Jun 2012 & 0.42 & 2.03--2.34 & 3000 & 3.63 \\
\noalign{\smallskip}
\hline
\end{tabular}
\end{table*}

For optimal subtraction of the atmospheric OH emission lines, we used nodding along the slit in an ABBA pattern, i.e. the star was observed before (A) and after (B) a first nod along the slit, then at position B a second time, before returning to position A for a final exposure. All stars were observed using two to five slit positions per cluster. The slit position angles were chosen such that two or more stars were observed simultaneously. The average signal-to-noise ratio (S/N) per pixel ranges from 50 to 150, and in a few cases S/N\,$\sim$\,30. Bright stars of spectral type B8 to A2 were observed as a measure of the atmospheric absorption and selected to share the same airmass as the targeted cluster stars during the middle of their observation.

\subsubsection{Reduction and extraction}

All reduction steps were executed using both custom-written Interactive Data Language (IDL) scripts and standard {\sc   iraf}\footnote{{\sc iraf} is distributed by the National Optical   Astronomy Observatories (NOAO), which is operated by the Association   of Universities for Research in Astronomy, Inc. (AURA) under   cooperative agreement with the U.S. National Science Foundation (NSF).} procedures.
%In the case of SofI spectra, we could first to correct for bad pixels, using the latest bad pixel mask available on the European Southern Observatory webpage\footnote{http://www.eso.org/lasilla/instruments/sofi/tools/\\reduction/bad\_pix.html}, and for cross-talk, as described in the SofI User's Manual (Moorwood et al. \cite{Mo98}). However, bad pixel masks and information about cross-talk were not available for ISAAC and OSIRIS, hence where omitted in the reduction procedure of these spectra. This is not be optimal, because cross-talk causes the signal of a bright star to be repeated somewhere else on the detector where another star can be present, and the bad pixels can introduce artificial structures at the scale of a few pixels. However, since the false signal potentially added to a given due to cross-talk is typically lower than 1\%, and considering that the first aim of the spectroscopic observations is spectral classification, both effects have a negligible impact on our final result.
Subsequent nodding observations were subtracted from one another to remove the bias level and sky emission lines. Flat fielding, spectrum extraction and wavelength calibration of all spectra were done in the usual way using {\sc iraf}. Calibration lamp spectra (helium-argon and neon for OSIRIS, xenon-neon for SofI and ISAAC) taken at the beginning of each night were used for wavelength calibration. The wavelength solution of each frame has an r.m.s. uncertainty between $\sim$0.3 and 0.5 pixels, which corresponds to $\sim$30~km~s$^{-1}$ for all instruments (except for ISAAC: see below). Correction for atmospheric absorption was done using the {\sc iraf} task {\it   telluric}. Before running this task, we first subtracted a fitted Voigt profile from the Br\,$\gamma$ absorption feature. Finally, all spectra were rectified using a low-order polynomial fit to a wavelength interval that was assumed to be pure continuum, i.e. where no absorption or emission lines were observed nor expected.

Unfortunately, the ISAAC pipeline could not be used, owing to many problems with the data. First, we had to drop the calibration lamp spectra, since \new{unrecoverable sudden} shifts along the spectral dispersion direction occurred during the night and a unique wavelength solution for the entire run could not be established. Instead, we used the OH sky lines for wavelength calibration. Since these lines were quite faint most of the time, owing to the short exposure times, the resulting calibration is not optimal and the r.m.s. uncertainty reached $\sim$0.8 pixels, i.e. $\sim$50~km~s$^{-1}$. We also had to take extra care of the ISAAC spectra because of severe, time-variable residual fringing. \new{The fringing pattern changes for every single frame and is different for the cluster stars and the standard star used for the correction of the atmospheric absorption. Hence it can not be removed by standard reduction and extraction}. The fringes are not constant along the spectra and hence could not be removed by simply subtracting a unique frequency from the Fourier transform of the spectra. On the other hand, the fringes in all spectra of all stars included in observations at a given slit position at a given time are very similar. Luckily, almost all the \new{stars} were observed twice during different nights, since the weather conditions during the first observing run did not meet our requirements (so they had to be repeated by the queue observers at Paranal). In these cases, the following procedure was adopted. From the 3--5 spectra per slit position, the spectrum of an OB star was selected. A Voigt profile was subtracted from the first observation (obs\,1) of this star at each \new{H\sc{i},} He{\sc i} and He{\sc ii} line (when present). The residual, dominated by the fringes, was then subtracted from the spectra of all other stars observed at the same time. These new, ``cleaned'' spectra were subtracted from the corresponding spectra of the other observation (obs\,2), and an average of the residuals was calculated. The averaged residuals were strongly dominated by the fringes of obs\,2 and included very little contribution from the individual stellar spectra. They could hence be used to cancel the fringing of all spectra of obs\,2. Once the spectra of obs\,2 had been cleaned, they were used to isolate the fringes in the spectra of obs\,1, which was, in turn, used to clean the spectra of obs\,1. This procedure was performed iteratively twice to ensure optimal removal of the fringes. Once the spectra of the two observations had been treated, they were averaged to obtain the final spectrum. Of course, this method of fringe removal is not perfect and may lead to residuals at the $\sim$5\% level, but this is sufficiently small so as to not prevent further analysis of the data.

\subsubsection{Spectral classification}\label{spcl}

The spectra of the stars in the fields of the six clusters are shown in Figure\,\ref{FigSp1}. Spectral classification was done based on comparison with available catalogues of $K$-band spectra of objects with spectral types derived from optical studies (Morris et al. \cite{Mo96}; Figer et al. \cite{Fi97}; Hanson et al. \cite{Ha96,Ha05}) as well as the spectral catalogs of Martins et al. (\cite{Ma07}), Crowther et al. (\cite{Cr06}), Liermann et al. (\cite{Li09}) and Mauerhan et al. (\cite{Ma11}). The most prominent lines covered by our wavelength range are Br\,$\gamma$ (4--7) 2.1661 $\mu$m (Brackett series), He{\sc i} lines at 2.1127 $\mu$m (3p$^3$P$^0$--4s$^3$S, triplet), He{\sc ii} \new{line} at 2.188 $\mu$m (7--10), the blend of C{\sc iii}\new{;} N{\sc iii} and O{\sc iii} at 2.115 $\mu$m, N{\sc iii}\,$\lambda$2.103 $\mu$m, and C{\sc iv} lines at 2.070 and 2.079 $\mu$m. All these lines were compared with the template spectra from these papers.

The equivalent widths (EWs) of the lines present in the spectral targets were measured on the continuum-normalized spectra using the {\sc iraf} task {\it splot}. In general, we used Gaussian profile fitting assuming a linear background. For more complicated profiles we used the deblending function or Voigt profile fitting. The EW uncertainties take into account the S/N of the spectra, the line's peak-to-continuum ratio (see Bik et al. \cite{Bi05}) and the error from telluric star subtraction (which was estimated at $\sim$10--15\% in the worst cases). The EWs of emission lines are negative by definition. Their values were used for qualitative estimation of the spectral types of OB stars using the calibrations given by Hanson et al. (\cite{Ha96}) and of WR stars using Figer et al. (\cite{Fi97}). In general, the spectra of most stars show \new{H{\sc i} and} He{\sc i} lines in absorption, which is typical for O type and early-B type MS stars. Also, 10 stars exhibit CO, Mg{\sc i}, Fe, Ca and Al{\sc i} lines, characteristic of late-type stars, and are classified as K- or M-type stars based on comparison with templates from the spectral library of Rayner et al. (\cite{Ra09}). Their luminosity class will be discussed in Section\,\ref{Results}. We discovered a total of 11 stars showing WR-like emission lines, i.e. 9 WN and 2 WC types. On the other hand, it is possible that some of these stars are OIf stars, since there is no clear way to distinguish them in near infrared $K$-band (Conti et al. \cite{Co95}). These stars will be discussed in more detail in Section\,\ref{newWR}. The method of spectral classification, especially in the near-IR, is correct within two subtypes, so that we adopt this error for our estimates. 

\new{When the S/N is sufficiently high, the luminosity class of OB stars is easily determined, since we can see if the Br\,$\gamma$ and He{\sc i} 2.16 line are blended (in MS stars) or not (in supergiant stars). However, for many of our spectroscopic targets, such features are not clearly seen, due to noise. For that reason, we verify if the luminosity class determined for a given star is coherent with that obtained for the other members of its host cluster. To do so, we measure the spectrophotometric distances obtained when luminosity classes V, III or I are assumed. In many cases, assuming that the cluster members are giants and/or supergiants yields a distance close to the edge of or outside the Milky way (Minniti et al. \cite{Mi11}), which makes this assumption unlikely. When all luminosity classes give plausible distances for the clusters, we use the distance to objects related to the clusters to take our final decision.} The estimated spectral types and luminosity class of all spectral targets are listed in Table\,\ref{TabSp}.

 \begin{table*}
\caption{Position, photometry and spectral parameters of the   spectroscopic targets.}
\label{TabSp}
\centering
\begin{tabular}{rrccrrrcccccc}
\hline\hline
\noalign{\smallskip}
&\multicolumn{1}{c}{ID}&      R.A.          &      Dec       &\multicolumn{1}{c}{$J$}&\multicolumn{1}{c}{$H$}&\multicolumn{1}{c}{$K_{\rm s}$} & {Sp. Typ.} & $E(J-K_{\rm s})$ & $E(H-K_{\rm s})$ &  $A_{K_{\rm s}}$ & $d$ & member\\
&            &   (J2000)     &   (J2000)    &                &              &                &                  &   (mag) &   (mag) &    (mag) & (kpc) & \\
\noalign{\smallskip}
\hline
\noalign{\smallskip}
\multicolumn{4}{l}{\it VVV\,CL009:} &&&&&&&\\
& 1 & 11:56:00.21 & --63:19:01.05 & 12.21 & 12.12 & 12.07 & B7--8V      & 0.18 & 0.05 & 0.12 & 2.\new{70} & \new{yes}\\
& 2 & 11:56:02.37 & --63:19:05.98 & 11.89 & 11.37 & 11.11 & O8--9V      & 0.9\new{6} & 0.\new{32} &  0.\new{63} & \new{5.52} & yes\\
& 3 & 11:56:02.55 & --63:19:05.19 & 12.25 & 11.70 & 11.47 & B1--2V        & 0.91 & 0.26 &  0.5\new{9} & \new{4.22} & yes\\
& 4 & 11:56:03.03 & --63:19:00.72 & 13.33 & 12.68 & 12.25 & B0--1Ve    & 1.21 & 0.47 &  0.7\new{9} & \new{5.53} & ?\\
& 5 & 11:56:03.07 & --63:18:54.63 & 11.87 & 11.31 & 11.04 & O8--9V      & 1.0\new{2} & 0.3\new{4} &  0.6\new{6} & \new{5.25} & yes\\
& 6 & 11:56:03.78 & --63:18:54.44 & 10.55 &   9.93 &   9.65 & WN7/Of     &          &          &   &  & yes\\
& 7 & 11:56:03.90 & --63:18:59.49 & 11.73 & 11.12 & 10.83 & O9--B0V    & 1.0\new{9} & 0.3\new{6} &  0.\new{71} & \new{4.67} & yes\\
& 8 & 11:56:04.38 & --63:18:54.44 & 12.14 & 11.59 & 11.34 & O9--B0V   & 0.95 & 0.29 &  0.\new{62} & \new{5.39} & yes\\
& 9 & 11:56:06.05 & --63:18:59.26 & 12.70 & 11.89 & 11.65 & K1--2V      & 0.39 & 0.13 &  0.2\new{5}	 & 0.26  & no\\
\multicolumn{13}{l}{}\\
\multicolumn{4}{l}{\it VVV\,CL011:} &&&&&&&\\
& 1 & 12:12:40.61 & --62:42:31.30 & 13.44 & 13.00 & 12.79 & B3/5V     & 0.74 & 0.23 &  0.4\new{8} & \new{4.65} & \new{yes}\\
& 2 & 12:12:41.13 & --62:42:30.71 & 11.72 & 10.60 & 10.05 & WN9 / OIf+       & 1.60 & 0.45 &  \new{1.04} & 10.\new{29} & yes\\
& 3 & 12:12:41.36 & --62:42:32.18 & 13.39 & 12.49 & 12.20 & B0--1V   & 1.33 & 0.32 &  0.8\new{7} & \new{5.20} & ?\\
& 4 & 12:12:41.40 & --62:42:29.86 & 13.81 & 12.77 & 12.24 & B1--2V   & 1.71 & 0.56 &  1.\new{11} & \new{4.73} & yes\\
\multicolumn{13}{l}{}\\
\multicolumn{4}{l}{\it VVV\,CL036:} &&&&&&&\\
& 1 & 14:09:00.49 & --61:15:54.01 & 15.28 & 12.45 & 11.06 & \new{B0--1V}     & 4.\new{41} & 1.4\new{6} &  2.\new{87} & \new{1.92} & yes\\
& 2 & 14:09:01.97 & --61:16:02.80 & 13.50 &   9.25 &    6.99 & M6--7I       & 5.56 & 2.04 & 3.\new{62} &   & yes\\
& 3 & 14:09:02.37 & --61:16:02.90 & 16.78 & 13.69 & 11.88 & B0--1        & 5.05 & 1.85 &  3.\new{29} & 2.\new{02} & no\\
& 4 & 14:09:02.75 & --61:15:58.94 & 14.25 & 10.95 &   9.32 & \new{B2--3}I    & 5.0\new{0} & 1.66 &  3.\new{25} & \new{2.60} & \new{yes}\\
& 5 & 14:09:02.83 & --61:15:33.48 & 15.53 & 12.65 & 11.24 & B0--1\new{V}    & 4.45 & 1.45 &  2.\new{90} & \new{1.80} & yes\\
& 6 & 14:09:03.28 & --61:16:05.87 & 15.33 & 12.65 & 11.33 &                   &         &          &   &  & ?\\
& 7 & 14:09:03.37 & --61:15:44.62 & 15.79 & 12.44 & 10.79 & \new{O8--9V}      & 5.\new{21} & 1.\new{75} &  3.\new{39} & 1.\new{46} & yes\\
& 8 & 14:09:04.23 & --61:16:06.59 & 12.17 & 11.13 & 10.56 & \new{O9--B0}V          & 1.7\new{6} & 0.6\new{1} &  1.\new{15} & \new{2.96} & \new{?}\\
& 9 & 14:09:04.30 & --61:15:53.78 & 13.37 & 10.66 &   9.16 & WN6         & 4.04 & 1.35 &  2.\new{63} & 2.\new{18} & yes\\
& 10 & 14:09:04.67 & --61:16:13.99 & 14.65 & 12.15 & 10.93 & \new{B0--1V}    & 3.\new{93} & 1.\new{32} &  2.\new{56} & 2.\new{29} & yes\\
\multicolumn{13}{l}{}\\
\multicolumn{4}{l}{\it VVV\,CL073:} &&&&&&&\\
& 1 & 16:30:21.57 & --48:12:55.84 & 13.05 & 11.76 & 11.27 & M7--8III  &          &          &   &  & no\\
& 2 & 16:30:23.73 & --48:13:04.90 & 15.40 & 12.51 & 11.08 & WN9 / O4--6If+  & 4.25 & 1.33 &  2.\new{76} & \new{7.47} & yes\\
& 3 & 16:30:23.92 & --48:13:24.68 & 12.51 & 11.41 & 10.96 & B0--1     & 1.71 & 0.49 &  1.\new{11} & \new{3.60} & ?\\
& 4 & 16:30:23.98 & --48:13:05.48 & 14.71 & 11.53 &   9.92 & WN7      & 4.70 & 1.53 &  \new{3.06} & \new{5.46} & yes\\
& 5 & 16:30:24.43 & --48:13:29.53 & 12.98 & 10.54 &   9.50 & K1--2III  &          &          &   &  & no\\
\multicolumn{13}{l}{}\\
\multicolumn{4}{l}{\it VVV\,CL074:} &&&&&&&\\
& 1 & 16:32:05.24 & --47:49:29.13 & 16.40 & 13.53 & 12.09 & \new{O8--9}III  & 4.\new{52} & 1.\new{54} &  2.\new{94} & \new{5.86} & yes\\
& 2 & 16:32:05.27 & --47:49:14.25 & 16.92 & 13.23 & 10.31 & WC8       & 6.18 & 2.53 &  \new{4.02} & 1.\new{54} & yes\\
& 3 & 16:32:05.46 & --47:49:28.10 & 14.72 & 11.82 & 10.17 & WN8       & 4.48 & 1.58 &  2.\new{9}2 & 2.\new{68} & yes\\
& 4 & 16:32:05.47 & --47:49:33.99 & 16.02 & 13.30 & 11.93 & B0--1      & 4.\new{30} & 1.4\new{7} &  2.\new{80} & \new{5.45} & yes\\
& 5 & 16:32:05.49 & --47:49:29.80 & 15.64 & 12.90 & 11.54 & \new{O7--8}III  & 4.\new{31} & 1.4\new{6} &  2.\new{81} & \new{4.84} & yes\\
& 6 & 16:32:05.55 & --47:49:33.19 & 16.00 & 13.24 & 11.89 & \new{O8--O9III}      & 4.2\new{3} & 1\new{41} &  2.\new{75} & \new{5.07} & yes\\
& 7 & 16:32:05.67 & --47:49:30.00 & 15.98 & 13.46 & 12.19 & \new{O7--8}III & \new{4.00} & 1.3\new{7} &  2.\new{60} & \new{7.16} & yes\\
& 8 & 16:32:05.86 & --47:49:30.48 & 16.65 & 13.35 & 11.73 & M2--3III   &          &          &   &  & no\\
& 9 & 16:32:05.93 & --47:49:30.92 & 15.22 & 12.53 & 11.31 & WN7 / O4--6If+   & 3.81 & 1.13 &  2.\new{48} & 1\new{3.57} & yes\\
& 10 & 16:32:06.43 & --47:49:26.46 & 16.33 & 12.74 & 11.11 &               &          &          &   &  & ?\\
& 11 & 16:32:06.99 & --47:49:21.05 & 16.30 & 13.05 & 11.58 & M1--2I  &          &          &   &  & yes\\
\multicolumn{13}{l}{}\\
\multicolumn{4}{l}{\it VVV\,CL099:} &&&&&&&\\
& 1 & 17:14:24.26 & --38:09:48.61 & 13.60 & 11.87 & 11.00 & O9\new{V}          & 2.7\new{8} & 0.9\new{4} &  1.\new{81} & 3.\new{13} & yes\\
& 2 & 17:14:24.41 & --38:09:50.76 & 14.72 & 13.11 & 12.31 & O9--B0V & 2.56 & 0.83 &  1.\new{66} & \new{5.80} & yes\\
& 3 & 17:14:24.50 & --38:09:54.59 & 14.38 & 12.97 & 12.38 & \new{B1--2}V      & 2.11 & 0.62 &  1.\new{37} & 3.\new{63} & no\\
& 4 & 17:14:24.71 & --38:09:49.35 & 12.62 & 10.93 &   9.91 & WN6+O   & 2.53 & 0.86 &  1.\new{65} & \new{4.84} & yes\\
& 5 & 17:14:25.42 & --38:09:50.40 & 13.43 & 11.70 & 10.64 & WN6        & 2.62 & 0.91 &  1.\new{70} & 6.\new{59} & yes\\
& 6 & 17:14:25.58 & --38:09:55.97 & 14.22 & 12.60 & 11.79 & B0/O9\new{V}      & 2.58 & 0.85 &  1.\new{68} & 4.\new{53} & yes\\
& 7 & 17:14:25.66 & --38:09:53.72 & 11.67 & 10.06 &   9.26 & WC8        & 1.98 & 0.42 &  1.2\new{9} & 3.\new{34} & yes\\
& 8 & 17:14:26.10 & --38:09:49.58 & 13.98 & 12.27 & 11.10 & O9\new{--B0V}           & 3.0\new{7} & 1.2\new{4} &  \new{2.00} & \new{2.99} & yes\\
& 9 & 17:14:26.34 & --38:09:43.81 & 10.91 &   9.01 &   8.26 & M4--5V      & & &   &  & no\\
& 10 & 17:14:26.42 & --38:09:49.87 & 15.21 & 12.96 & 12.02 & M1--2V   & & &   &  & no\\
& 11 & 17:14:26.45 & --38:09:53.23 & 15.07 & 13.03 & 12.16 & M1--2V   & & &   &  & no\\
& 12 & 17:14:26.68 & --38:09:45.05 & 14.12 & 12.41 & 11.51 & M4--5V   & & &   &  & no\\
\noalign{\smallskip}
\hline
\end{tabular}
\tablefoot{ Columns include the name of the star, its   position (R.A. and Dec), $J$, $H$ and $K_{\rm s}$ photometry (in   mag), spectral type, $E(J-K)$ and $E(H-K)$ colour excesses (in mag),   extinction ($A_K$) and distance ($d$), respectively. The final column provides our diagnostic with regards to cluster membership (or otherwise).}
\end{table*}

\section{The method}\label{Met}

Our methods to determine the fundamental parameters of the massive clusters, such as their angular sizes, extinction values, distances, ages and masses, have already been described in Paper~I. Here, we provide a brief summary and highlight what we have done differently in this study.

\subsection{Angular size}\label{AngSiz}

\begin{figure}
  \centering
   \includegraphics[width=9.cm]{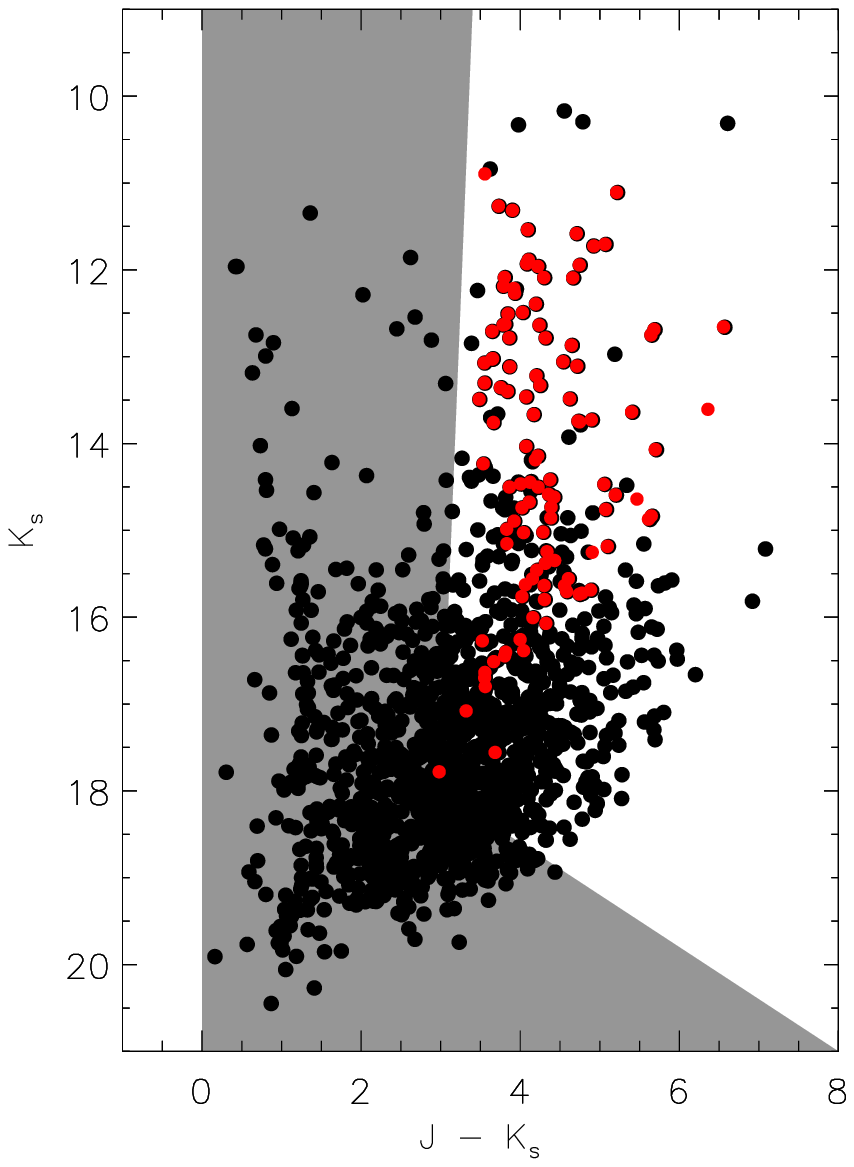}
   \caption{$(J-K_{\rm s})$ colour as a function of $K_{\rm s}$ for VVV\,CL074 based on the complete photometric catalogue (black points). Only stars within a radius of 0.7$'$ around the central position of the cluster are plotted. This radius was derived from the RDP calculated from the complete catalogue. The red dots represent the same CMD, but based only on the decontaminated catalogue. The stars in the gray area are undoubtedly not part of the cluster and can be removed from the catalogue without any loss of information before recalculating the RDP. Once a new and refined cluster radius has been determined from the RDP obtained from the filtered catalogue, a final statistical decontamination step was performed (see text).}
              \label{filt}%
\end{figure}

\begin{figure}
  \centering
   \includegraphics[width=9.cm]{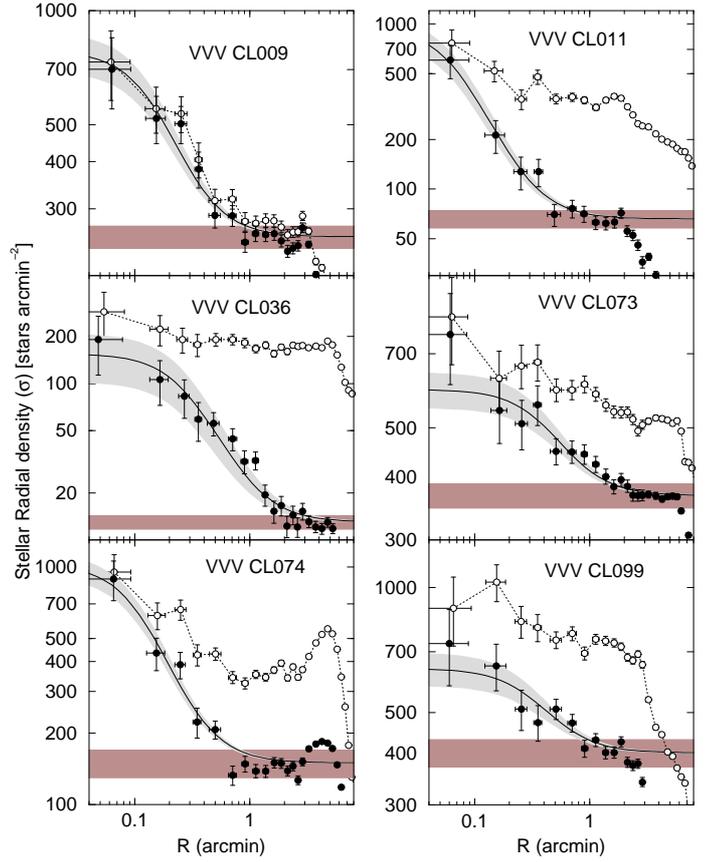}
   \caption{RDPs of the six clusters. The open circles show the RDPs obtained from the complete catalogue, while the black points are based on the filtered catalogue. The red horizontal band represents the background level (the width is determined by the uncertainty on the value) and King profiles are given with uncertainties in gray.}
              \label{RDP}%
\end{figure}

We first obtained the coordinates of the cluster centre based on our stellar catalogue. Starting from a first-guess value, we calculated the radial density profile (RDP) iteratively until it converged towards the best profile (i.e. with the highest and clearest difference between the background and the central density levels). The central coordinates of the clusters are given in Table\,\ref{TabCl}. Once the converged RDP has been obtained, we evaluate the density of field stars in the vicinity of the cluster, which we define as the background. Since we are dealing with lower-density clusters than in Paper\,I, some RDPs may not be very clear. Hence, we systematically removed all stars from the catalogue that are obviously field stars. To do so, we localized the position of the cluster in the CMD using a decontaminated catalogue, assuming a cluster radius obtained from the RDP calculated using the full catalogue. Next, we assumed that, for each magnitude bin, all stars bluer than the bluest cluster stars must originate in the field (see example in Figure\,\ref{filt}) and can be removed. This technique is optimal for highly reddened clusters and leads to higher-contrast RDPs. The RDPs of five of the six clusters discussed in this study were thus significantly improved, while the RDP of the sixth cluster benefited only mildly. All cluster RDPs are shown in Figure\,\ref{RDP} and will be discussed individually in Section\,\ref{Results}.

To better guide our determination of the clusters' angular sizes, we fit a two-parameter King profile adapted to star counts (as in King \cite{Ki66}, but using star counts instead of surface brightness). The two free parameters are the projected central stellar density ($S_0$) and the core radius ($r_{\rm{c}}$). The uncertainties on the fit are given by the maximum and minimum values allowed within the observational error bars (given by Poissonian noise). The angular size of a cluster is defined as the radius where the RDP meets the level of the field stars.
%Up to here.
\subsection{Cluster membership}\label{mem}

\new{The present statistical decontamination approach is far from perfect with respect to the determination of cluster membership. Indeed, the algorithm only considers the number of stars that are in excess (computed for a discrete distribution of colour and magnitude cells) within the cluster field with respect to the comparison field. Since the decontamination algorithm does not take stellar parameters into account (not even the reddening calculated based on the CCD) it arbitrarily selects which stars should be removed or retained. If a field star is kept at the expense of a real cluster member, it does not alter the diagnostic obtained from the CMD itself, since both stars have comparable colors and magnitudes. The ideal way to determine membership is to measure the radial velocities (RVs) and the the proper motions of the stars in the clusters region, similarly to what we presented in Paper\,I. 

Unfortunately, the VVV database currently contains observations covering only two years, which is insufficient to determine precise proper motions. It was not possible to construct proper motion histograms using an absolute proper motions catalogue (as done in Paper\,I), since the large extinction, small size and, in some cases, the compactness of the clusters precluded them from having been appropriately observed in optical low-resolution surveys. Indeed, based on a search of the PPMXL Catalog of Positions and Proper Motions on the ICRS (Roeser et al. \cite{Ro10}), only three stars could be matched in VVV\,CL009, two in VVV\,CL011, one in VVV\,CL036, and none in each of VVV\,CL073, VVV\,CL074 and VVV\,CL099. 

RVs were obtained for our  spectroscopic targets. Unfortunately, the low resolution of the spectra, combined with the small wavelength coverage, does not allow a very accurate measurement of the RVs. Moreover, considering that the great majority of the spectra suffer from a poor S/N and/or fringing and/or artificial shifts in the wavelength solutions, it was impossible to obtain any usable measure for more than one or two stars per cluster. This is insufficient to provide any diagnostic as to a cluster's RV or to verify the membership probability of any of the observed stars.}

Instead, we checked the loci of the stars in the $(J-K_{\rm s})$ vs. $K_{\rm s}$ CMD (see Figure\,\ref{CMD}). Of course, the results obtained this way can be affected by differential reddening. We attempted to reduce the effects of differential reddening by employing the reddening-free parameter \new{$Q_{\rm NIR} =(J-H)-\alpha(H-K_{\rm s})$, where the extinction power law $\alpha=2.14$, as determined by Stead \& Hoare (\cite{St09};} see also Catelan et al. \cite{Ca11} for a list of several other reddening-free indices in the $ZYJHK_{\rm s}$ VISTA system). We chose this parameter to avoid the intrinsic degeneracy between reddening and spectral type (and since we expect to find OB stars in the clusters of our sample). Originally, Negueruela et al. (\cite{Ne11}) defined $0.0 \leq Q_{\rm NIR} \leq 0.1$\,mag as a separating value for early-type stars. Subsequently, Ram{\'{\i}}rez Alegr{\'{\i}}a et al. (\cite{Ra12}) and Borissova et al. (\cite{Bo12}) defined the separation value for OB stars as $0.3 \leq Q_{\rm NIR} \leq 0.4$\,mag. Here, we prefer to plot MS isochrones (using red dashed lines) for an age of 5\,Myr (i.e. close to the average of all cluster ages and young enough to clearly show the MS up to the highest masses). \new{Interestingly, the choice of $\alpha=2.14$ puts most of the MS stars close to the value $Q_{\rm NIR}=0$.}

The reddening-free CMDs (see Figure\,\ref{CMD}) can also be used to separate MS from pre-MS (PMS) stars (gray points), information which was used for our isochrone fitting and mass calculation (see below). Special care has been taken in relation to the cool stars in the clusters, since they could be RSGs, foreground red-clump giants or late-type MS stars. In addition to their position in the $Q_{\rm NIR}$ diagram, we plot the EW of the CO line, measured between 2.294 and 2.304 $\mu$m, as a function of spectral type. If the EW(CO) is not consistent with the relations of Negueruela et al. (\cite{Ne10}) for giant and supergiant stars, we do not consider the star a cluster member. 
%[TO ADD, MIR PHOTOMETRY FOR INVESTIGATING COOL STARS]

\new{We remark that, since we still do not provide an unambiguous membership diagnostic, we can not conclude wether a given star (projected towards a cluster) is a real member of its cluster or not. Here, we define a uniform approach to select the likely member stars used to calculate the clusters parameters.}

\subsection{Extinction, spectrophotometric distance and reddening}

Our spectroscopic classification and IR observations permit direct evaluation of the clusters' extinction, distance and applicable reddening laws. Intrinsic $JHK_{\rm s}$ colours and absolute magnitudes were adopted from Schmidt-Kaler (\cite{Sc82}) \new{for B stars and Martins et al. (\cite{Ma07}) for O stars}. Dust extinction at $K_{\rm s}$ was evaluated using $A_{K_{\rm s}}=0.6\times E(J-K_{\rm s})$, and the distance moduli follow: $\mu_0=K_{\rm s}-M_{K_{\rm s}}-A_{K_{\rm s}}$. The standard deviation and standard error were computed, and cases where the former is large may be indicative of differential reddening.

Decontaminated $(J-H)$ vs. $(H-K_{\rm s})$ colour--colour diagrams (CCD) are presented in Figure~\ref{CMD}. O and B stars are marked with blue and red points, respectively. The intrinsic colours of MS and giant-branch stars (Koornneef \cite{Ko83}) are plotted using blue dashed lines. The reddening vector, plotted in orange, was determined from the calculated spectroscopic reddening. Colour-excess ratios, e.g., $E(J-H)/E(H-K_{\rm s})$, were tabulated for each cluster and \new{to within the uncertainties the results generally match recent estimates of the reddening law as inferred from numerous sight-lines towards the inner Galaxy (Strai\^zys \& Lazauskait\`e  \cite{St08}, Majaess et al. \cite{Maj11}, \cite{Maj12})}, except for one cluster. Sources located to the right and below the reddening line may exhibit excess emission in the near-IR (IR-excess sources) and/or be PMS stars.

Spectrophotometric distances to the stars for which we obtained spectra were calculated: see Table\,\ref{TabSp}. For each cluster, we selected the stellar members for which we are confident of their spectral types. The resulting error \new{is obtained using propagations of uncertainties of all the values involved in the calculation of the distance}.

\subsection{Ages}

The age can be estimated by fitting the observed CMD with non-rotating Geneva isochrones (Lejeune \& Schaerer \cite{Le01}), combined with PMS isochrones (Seiss et al. \cite{Se00}). The Geneva isochrones are the best publicly available models for stars with masses greater than 10\,$M_\odot$, but they are only available for solar metallicity. We have no direct way to determine the metallicity of our sample clusters, and we cannot evaluate how far from reality assuming solar values may be. On the other hand, \new{the clusters are relatively close to the Sun} (see Table\,\ref{results}), so solar metallicity is likely a reasonable assumption. 
%[VERIFICATION USING   PADOVA?]

Starting from the isochrones scaled to the previously determined distance modulus and reddening, we apply shifts in magnitude and colour until the goodness-of-fit statistics reach a minimum value (i.e. the differences in magnitude and colour of the observed stars from the isochrone should be minimal) for both the MS and PMS isochrones. For such young star clusters (i.e. $<$ 10 Myr) the MS isochrones are almost vertical lines in the near-IR and it is hard to determine the precise age, even when using the PMS isochrone set. For this reason, we determine an interval of possible ages based on the range of isochrones that could be fitted satisfactorily. The results are shown in Figure\,\ref{CMD}, where the PMS stars are plotted as gray points. Modelling of the spectra using {\sc cmfgen} (Hillier \& Miller, 1998) is currently being done and will be presented in a forthcoming paper (A.-N. Chen\'e et al., in prep). Once the stellar fundamental parameters of the cluster members have been determined precisely using these models, they will be used to construct HR diagrams, allowing us to compare the distribution of these stars with the latest isochrones and stellar tracks (including rotation) of Ekstr{\"o}m et al. (\cite{Ek12}).

\subsection{Present day mass function and cluster masses}

We first used the cluster CMDs to select only cluster members on the MS. All rejected stars are either byproducts of our decontamination method (see section \ref{obs_phot}) or evolved (WR, RSG) stars. The luminosity function (obtained as described in Section\,\ref{obs_phot}) was converted to the present day mass function (PDMF) using the Geneva isochrones, assuming solar metallicity and using the intermediate ages of the clusters determined as a average value of the age intervals as described above. For each cluster we established the MS turn-on point and assumed that all fainter objects were PMS stars. Figure \ref{FigMF1} shows the PDMF of the clusters. The mass range over which the slope is calculated is shown as vertical dashed lines: only MS objects were considered. An upper mass limit was set to where the PDMF slope starts to skew due to the effects of saturated stars, with the lower limit set to where the effects of incompleteness affect the slope. Uncertainties are also shown; these were calculated from the propagation of small-number stochastical uncertainties from Gehrels (\cite{Ge86}). Note that despite the fact that we refer to the cluster PDMFs and despite the young age of the clusters, the stellar mass function cannot be considered the true PDMF. Dynamical cluster evolution will result in stellar losses through energy equipartition, resulting in a higher average stellar mass and an increase in the binary fraction. This will affect the slope of the stellar mass function (causing it to flatten). This is sensitive to both mass and cluster lifetime (in particular for these clusters, which have masses of $<$10$^4$\,M$_{\odot}$ and lifetimes of a few 100\,Myr; Kroupa et al. \cite{Kr11}) so this effect must be duly considered for these clusters. In addition, due to stellar evolution the number of observable massive stars with initial masses equal to or greater than 60\,$M_\odot$ decreases with time, and clusters with an age of 5\,Myr can already have a mass function showing an abrupt drop at high masses. Our calculated slopes $\Gamma$ are listed in Table\,\ref{results}, with uncertainties that propagate from those on the individual mass bins. These slopes were used to extrapolate the masses of the observed main sequence stars to provide a total masses for the clusters. The large uncertainties on the mass are due to the small mass-sampling region of the PDMF slope. As a result the total masses that we have obtained should be treated as a lower limit and taken cautiously until deeper photometry can be analyzed.

\section{Results}\label{Results}

In this section, the six clusters will be discussed in more detail. Since all objects have one or more WR stars in their central regions, we consider them real clusters \new{in this section, but the likelihood that they could not be} will be discussed in Section\,\ref{realclust}.

\subsection{VVV\,CL009}

The RDP of VVV\,CL009 shows clear evidence of the presence a cluster, based on both the complete and the filtered catalogue (open and filled circles, respectively). The central position of the cluster associated with the optimal RDP is R.A. (J2000)\,=\,11:56:03.18 and Dec (J2000)\,=\,$-63$:18:42.19. Our fit of a two-parameter King profile to the RDP reveals a background density ($S_{\rm B}$) level at $253\pm17$\,stars arcmin$^{-2}$, while $S_0=530\pm90$\,stars arcmin$^{-2}$ and $r_{\rm{c}}=0.17'\pm0.03'$. The cluster radius reaches $0.6'\pm0.2'$ (equivalent to $1.1\pm0.5$\,pc using the distance determined below).

Upon examination of the stars' positions in the CMD, we see that star 1 is located quite far off towards the blue. \new{However, it seems to be the effect of some differential reddening, since in the plot showing the reddening-free parameter, this same star is found along the MS. Hence, we assume that this star is a cluster member.} As for star 4, it exhibits emission lines, but the low S/N of its spectrum does not enable us to determine its spectral type with confidence. Based on the strength of its Br\,$\gamma$ line in emission, we identify it as a B0--1 Ve star, but we cannot distinguish it from a young stellar object. Hence we will not use it for the determination of the cluster parameters. The spectral classification of star 9 gives K1--2V. Its measured EW(CO) is 5.5\,\AA, which corresponds to a G2--5 spectral type if the star were a giant or supergiant. \new{Hence, this star is likely not a member of the cluster.}%Moreover, [SOMETHING ABOUT THE MIR PHOTOMETRY].

Using stars 2, 3, 5, 7 and 8 (star 6 is omitted since uncertainties in the absolute magnitudes of WR stars are very large \new{and star 1 was excluded to the determination of the $A_K$-value, due to a different reddening from the rest of the cluster members, but included in the determination of the distance}), we obtain average values of $A_K=0.64\pm0.01$\,mag ($A_V=5.4\pm0.1$\,mag) and a distance $d=5\pm1$\,kpc. The ratio $E(J-H)/E(H-K_{\rm s})=2.11\pm0.05$, which seems to be in little excess of the values for the inner Galaxy listed in Table\,1 of Strai\^zys \& Lazauskait\`e (2008). This would be the only cluster in this study exhibiting an anomalous reddening law along the line of sight. This result is analogous to that obtained for Danks\,2 in Paper\,I. Some PMS stars can be distinguished and are defined using $Q_{\rm NIR}=0.30$\,mag and $K_{\rm s}=16.5$\,mag as diagnostic cuts. MS stars fall in the range $-0.1 \leq Q_{\rm NIR} \leq 0.1$\,mag. Other stars, including star 9, are found in $0.3 \leq Q_{\rm NIR} \leq 0.5$\,mag and could be background stars that were not properly decontaminated, or even some contribution of a less reddened cluster along the same line of sight. Given the distance and reddening, the fit using PMS isochrones suggests an age near or greater than 5\,Myr, while the range of ages supported by the MS isochrones is 4--6\,Myr, which is perfectly compatible. The slope of VVV\,CL009 PDMF is $\Gamma =-0.7\pm0.2$ and\new{, using 111 MS stars, we obtain an observed mass of $660\pm150$\,$M_\odot$. The lower limit for the total extrapolated mass is 1000\,$M_\odot$.}

\subsection{VVV\,CL011}

Based on the complete catalogue resulting from the VVV-SkZ\_pipeline, this cluster's RDP shows a high background level and some over-density of stars centered on R.A. (J2000)\,=\,12:12:41.22 and Dec (J2000)\,=\,$-62$:42:33.52. The profile becomes much clearer once the decontaminated catalogue is used leading to a much lower background level and a much steeper profile. Note that the sudden decrease in the background level beyond a radius of $\sim 3'$ is due to the edge of our photometric catalogue, hence we simply exclude this section of the RDP from our analysis. The parameters of the best-fitting King profile are $S_0=900\pm171$\,stars arcmin$^{-2}$ and $r_{\rm{c}}=0.07'\pm0.01'$, and we also derive $S_{\rm B}=66\pm8$\,stars arcmin$^{-2}$. The cluster radius is $0.5'\pm0.1'$ (i.e. $\sim$0.9\,pc).

Looking at the CMD, we see that star 1 occupies a position that seems to be off towards the blue. \new{On the other hand, the $Q_{NIR}$ parameter places it among the MS stars, and we consider it as a cluster member. Star 3 was removed from the photometric catalogue by the decontamination algorithm without leaving behind any other star close to its positions.} This is an indication (but not proof) that the section of CMD space occupied by star 3 should not contain any cluster members. \new{Moreover, the low S/N of star 3's spectrum does not allow any accurate spectral classification and its membership can not be clearly determined. This leaves only stars 1 and 4 for calculation of the parameters of VVV\,CL011,  since we prefer to avoid using the WR star 2. Also, star 1 was excluded from the calculation of the reddening, since it likely suffers from local differential reddening and its inclusion would move all isochrones towards the blue despite most of the other points being redder.} This yields $A_K\sim1.11\pm0.02$\,mag (i.e. $A_V\sim9.4\pm0.1$\,mag) and $d=5\pm2$\,kpc. We identify very few PMS stars; however, if we assume that the MS is located along $Q_{\rm NIR}=0.0$\,mag, many more stars could be considered PMS stars. We hence are left with very poor constraints on the age of this cluster because of the small number of data points, and the cluster could either be as young as 3\,Myr or as old as 7\,Myr. \new{The observed mass obtained using 67 MS stars is $390\pm150$\,$M_\odot$, and the lower limit for the total extrapolated mass is 950\,$M_\odot$ when $\Gamma =-1.0\pm0.6$ is used.}

VVV\,CL011 is a quite small and compact cluster which may need to be revisited using an integral field spectrograph to observe more stars than can be achieved using long-slit spectroscopy. Also, we note that the spectrum of VVV\,CL011-2, which we classified WN9, is also compatible with an OIf type. Later analysis will allow us to determine the real spectral type of this star (A.-N. Chen\'e et al., in prep). \new{On the other hand, we can already see that the distance to the star obtained is $8.0\pm0.2$\,kpc when a OI spectral type is used, which is significantly farther than the distance determined for the cluster.} 

\begin{figure*}
  \centering
   \includegraphics[width=14.cm]{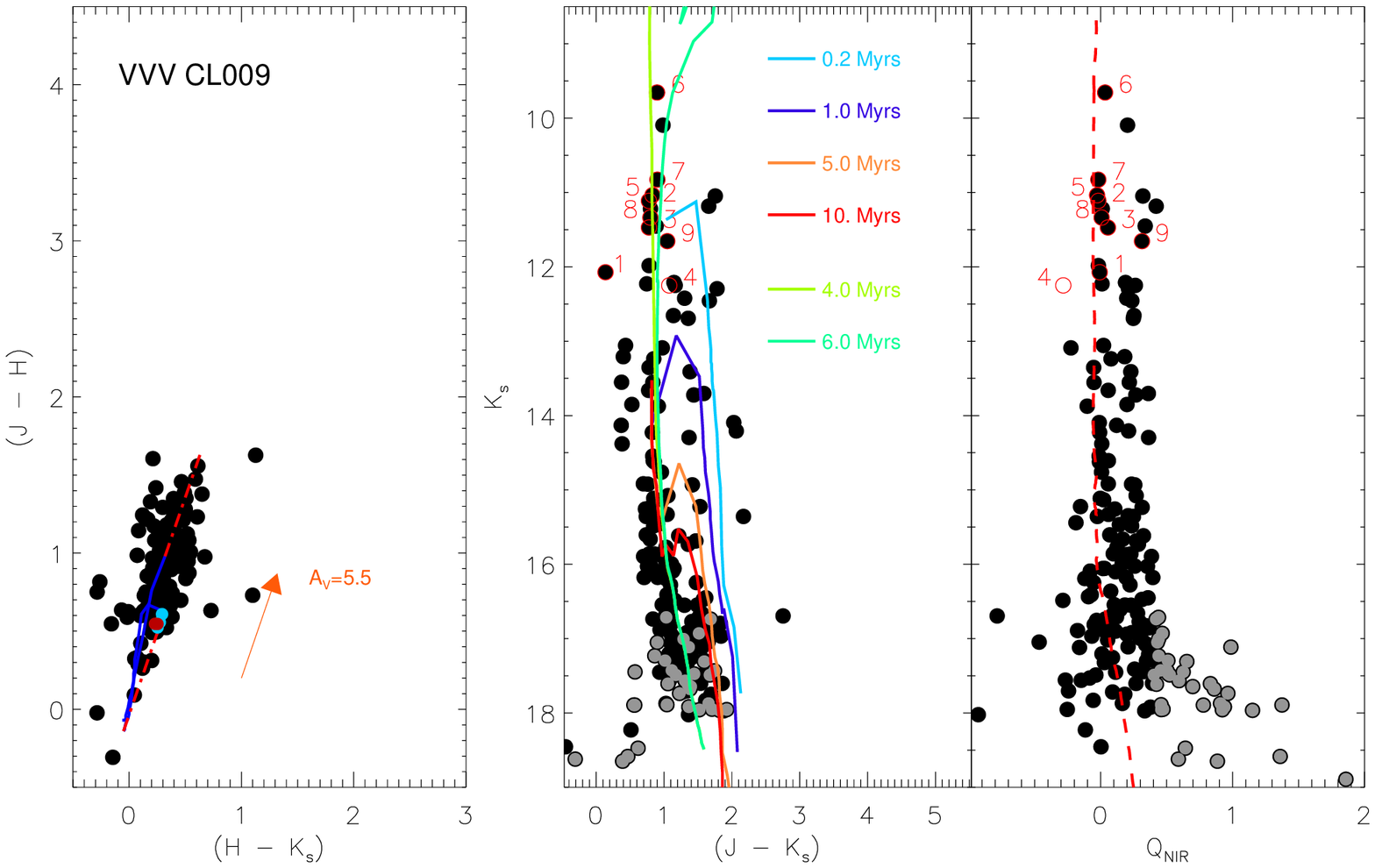}
   \includegraphics[width=14.cm]{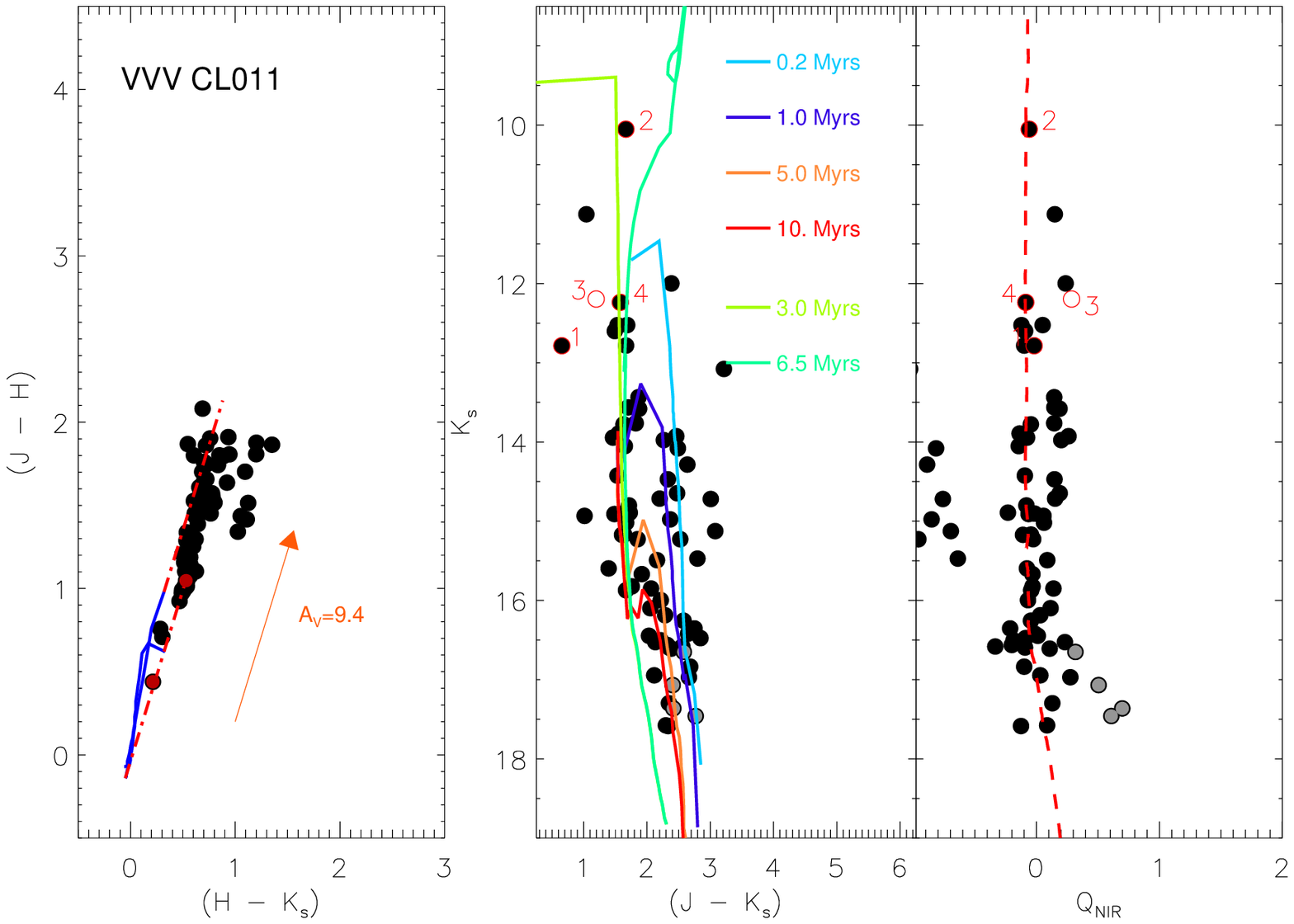}
   \caption{For each cluster we present a series of three diagrams. {\it Left}: $(H-K_{\rm s})$ vs $(J-H)$. The Koornneef et al. (\cite{Ko83}) and Schmidt-Kaler (\cite{Sc82}) models for luminosity classes I and V, respectively, are plotted in blue. The O and B stars are plotted as small blue and red points, respectively and the reddening vector indicates the measured value of $A_V$. {\it Middle}: $(J-K_{\rm s})$ vs $K_{\rm s}$. Spectroscopic targets are marked using red circles, but those with magnitudes or colours outside the figure boundaries are listed on the plot. The PMS isochrones are shown in light blue (0.2\,Myr), dark blue (1.0\,Myr), orange (5.0\,Myr) and red (10\,Myr), while the two upper and lower limits of fitted MS isochrones are shown in light and dark green. {\it Right}: $Q_{\rm NIR}=(J-H)-2.14(H-K_{\rm s})$ vs $K_{\rm s}$. Spectroscopic targets are also marked. A 5\,Myr MS isochrone is plotted using a dashed red line and the PMS stars are plotted as gray points.}
              \label{CMD}%
\end{figure*}
\setcounter{figure}{4}
\begin{figure*}
  \centering
   \includegraphics[width=14.cm]{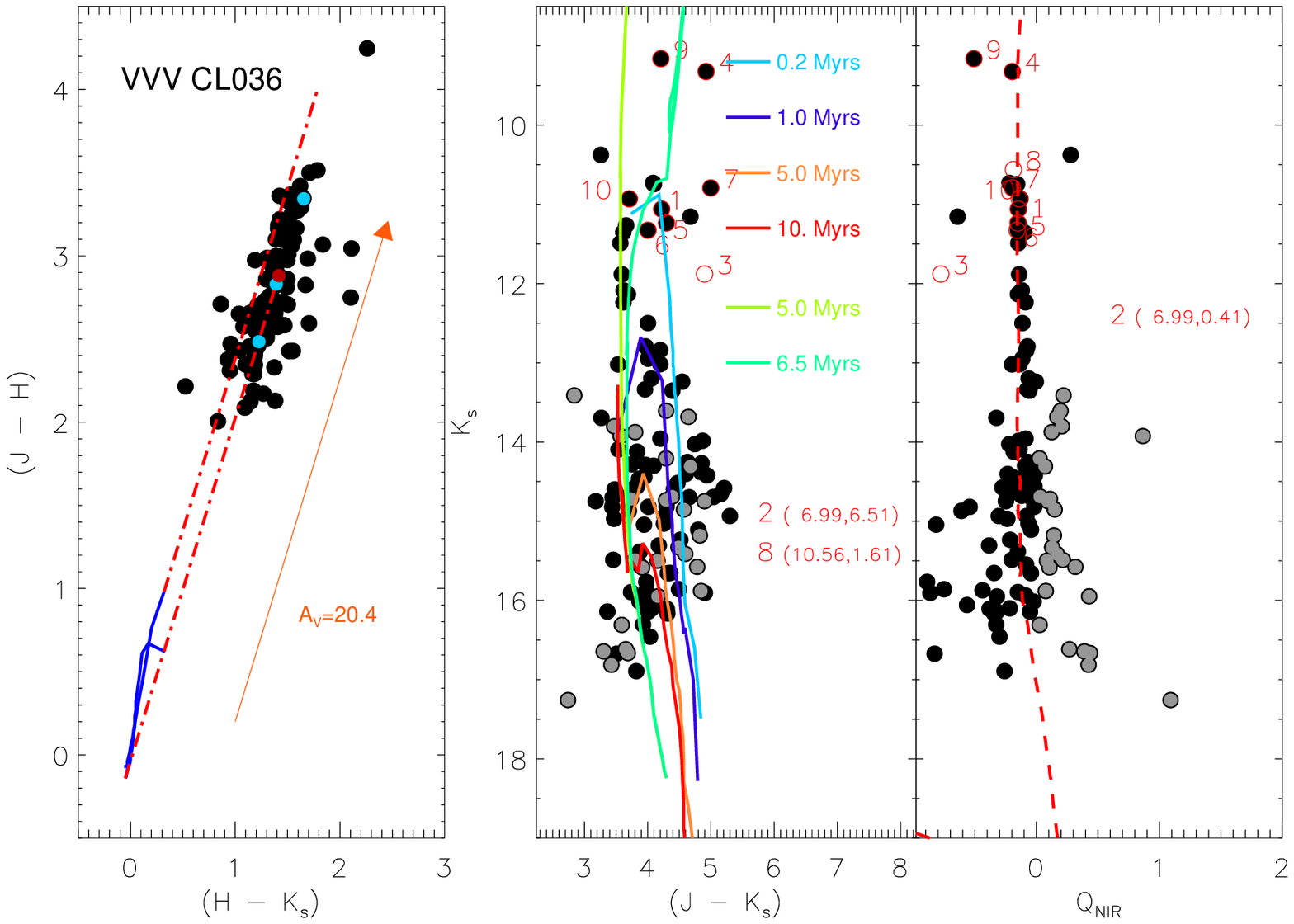}
   \includegraphics[width=14.cm]{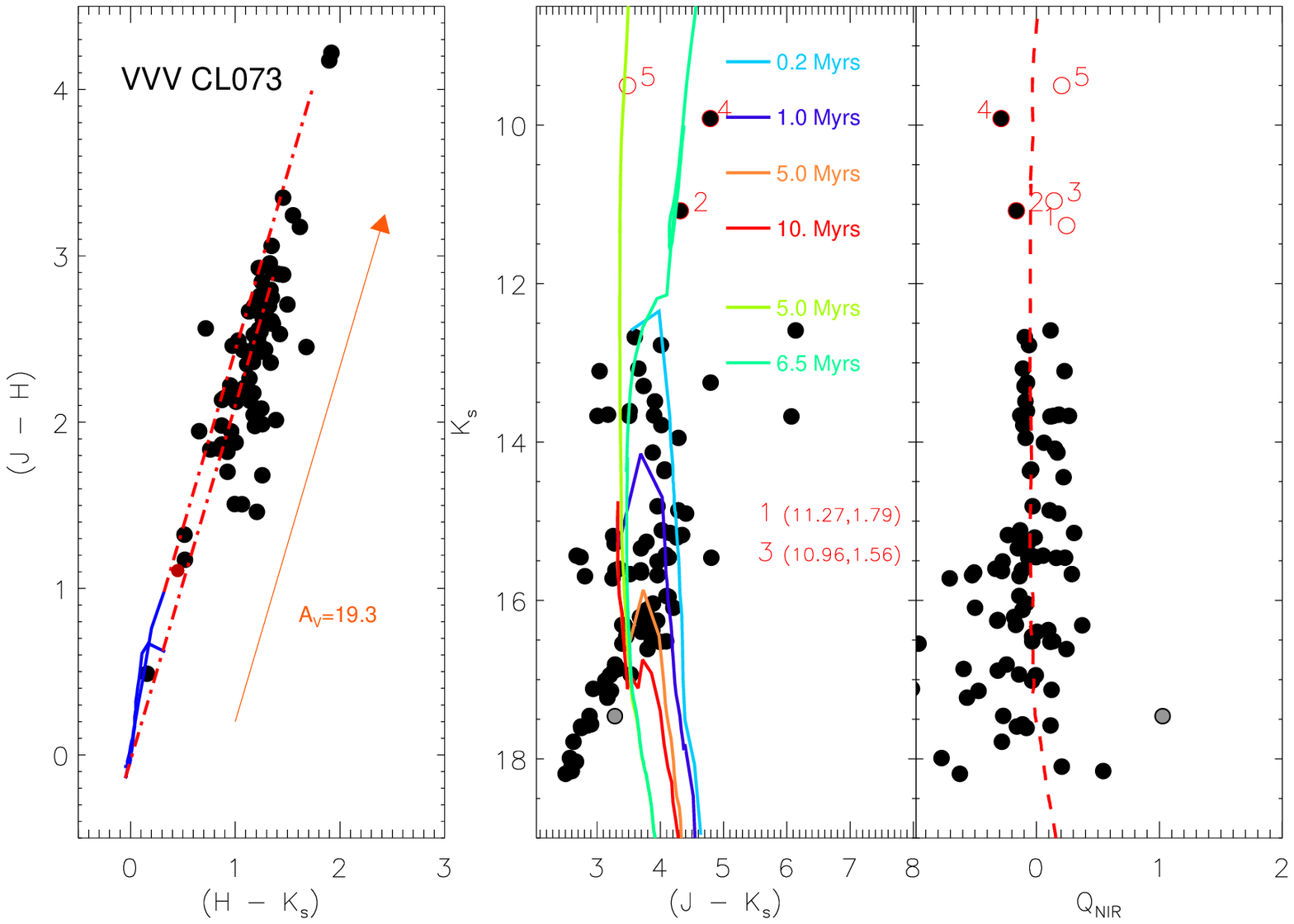}
   \caption{(continued)}
\end{figure*}
\setcounter{figure}{4}
\begin{figure*}
  \centering
   \includegraphics[width=14.cm]{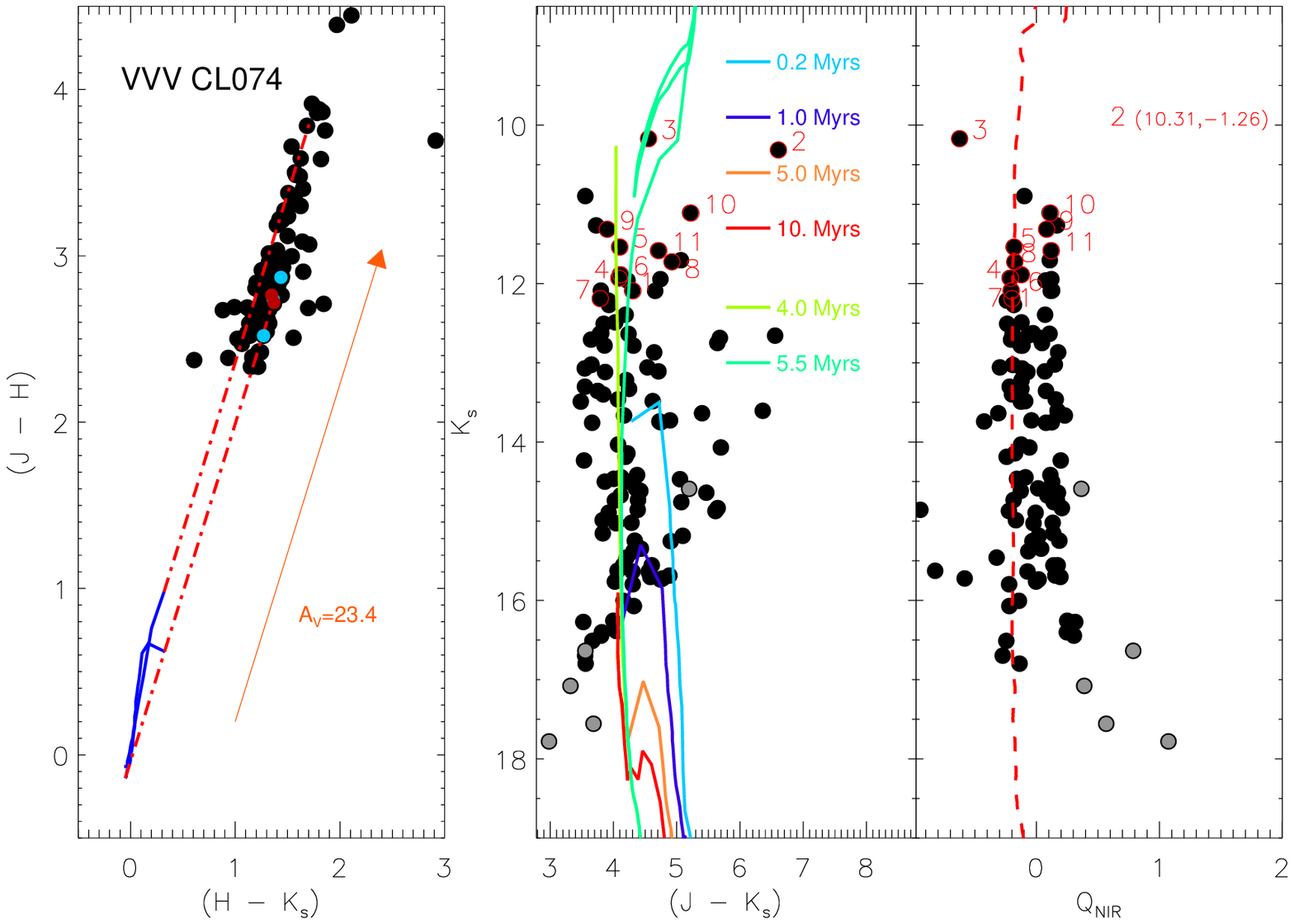}
   \includegraphics[width=14.cm]{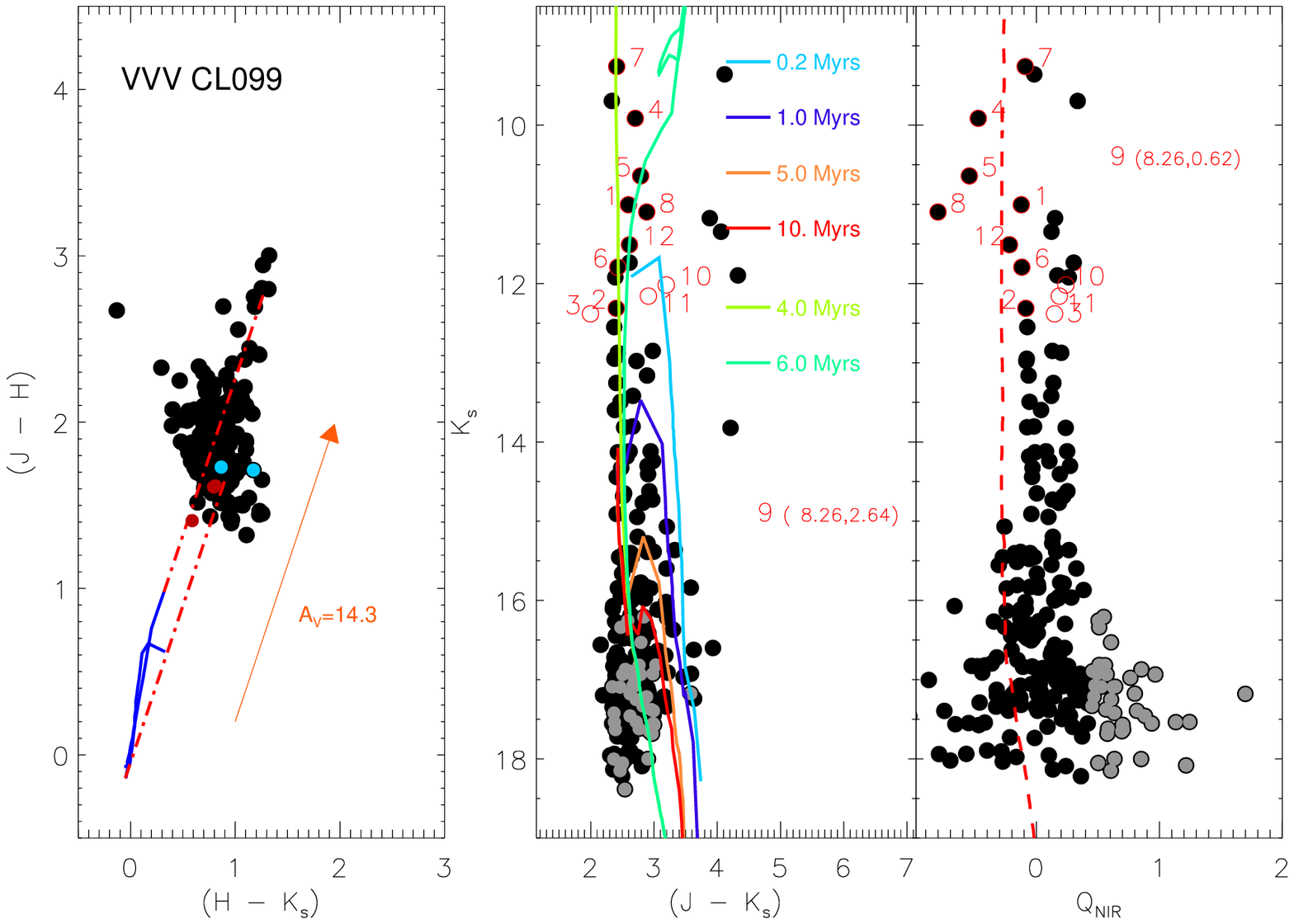}
   \caption{(continued)}
\end{figure*}

\subsection{VVV\,CL036}

This cluster's RDP based on the complete catalogue shows very little evidence for the presence of a cluster. Fortunately, this cluster is quite heavily reddened (see below) and the filtering of the field stars was done very efficiently. The cluster centre is located at R.A. (J2000)\,=\,14:09:03.95 and Dec (J2000)\,=\,$-61$:15:58.14. The parameters of the best-fitting King profile are $S_0=142\pm52$\,stars arcmin$^{-2}$ and $r_{\rm{c}}=0.29'\pm0.07'$, and we also find $S_{\rm B}=13\pm1$\,stars arcmin$^{-2}$. The cluster extends beyond a radius of $1'$, but the angular distribution is very patchy and our analysis becomes meaningless at larger radii. Hence, we adopt a cluster radius of around $1'$ (i.e. $\sim$0.8\,pc).

Stars 3 and 8 do not seem to be related to the other stars in the CMD. Star 3 is rather faint for its spectral type and red, and star 8 is very blue. \new{However, the Q$_{NIR}$-value for star 8 places it among the MS stars. Therefore, we consider star 3 as a field star, but give no definitive diagnostic on the membership of star 8}. As the S/N of their spectra is quite low ($\sim$30), the uncertainty associated with the spectral classification is larger than for our other spectroscopic targets, and they may be cooler than B stars. Star 2 is a very bright M6--7I star. Its $K_{\rm s}$ magnitude places it outside the boundary of our CMD. It EW(CO) value is 33 \AA, which is compatible with its spectral classification. 
%[I EXPECT THE   MIR PHOTOMETRY TO CLEARLY INDICATE THIS STAR IS A RSG]. 
Considering this, we assume this star to be a cluster member. If confirmed it will be interesting case where RSG and WR stars are formed simultaneously in intermediate massive cluster.\new{Star 4 is a bright (nearly as bright as the WR star) B supergiant.} The spectrum of star 6 has a S/N that prevents its use here.

The OB stars are spread over a wide range of reddening in the CCD. This indicates that the cluster suffers from significant differential reddening. Thus, the average value from the six remaining stellar members of VVV\,CL36, after excluding the WR star, is not sufficient to obtain a good determination of the cluster parameters. We therefore use only the CCD and CMD to get an average value for the reddening, and we use the scatter we find in our spectroscopic values as an indication of the uncertainty in the differential reddening. This yields $A_K=3.0\pm0.3$\,mag ($A_V=24\pm3$). \new{As for the spectrophotometric distance, we get $2\pm1$\,kpc. Some H{\sc ii} regions and YSOs with measured RVs are found near the cluster (Urquhart et al. \cite{Ur07,Ur09}). On average, we get a kinematic distance to the system of 4.3\,kpc, which is in agreement with our spectroscopic distance.} PMS stars cannot be distinguished from the other stars. Here, we assume $Q_{\rm NIR}=0.0$\,mag as the boundary, but due to the differential reddening the fit of the PMS isochrones does not yield any relevant result. On the other hand, MS stars could be quite well isolated using $-0.25 \leq Q_{\rm NIR} \leq -0.05$\,mag, and the isochrone fit yields an age range of 5--6.5\,Myr. The PDMF slope is $\Gamma=-1.0\pm0,4$ and leads to \new{a lower limit for the total extrapolated mass of 2200\,$M_\odot$ (observed mass obtained with 207 MS stars is $1800\pm300$\,$M_\odot$).}

\subsection{VVV\,CL073}

The unfiltered RDP exhibits two plateaus: one that results in a cluster radius of around $0.5'$ (i.e. $\sim$0.6\,pc) and another of around $1.0'$ (i.e. $\sim$1.2\,pc). Considering the nature of the complex background (see Figure\,\ref{FOV}), which contains obscured clouds, these measurements of the cluster radius may be biased. For this cluster, we obtain R.A. (2000)\,=\,16:30:24.20 and Dec (2000)\,=\,$-48$:13:00.44. The only two cluster members that could be observed spectroscopically are the WR stars VVV\,CL073-2 and 4. The other spectroscopic targets are either near or outside the cluster's radius. Moreover, stars 1 and 3 are too blue to be cluster members, and the EW(CO) value (18.9\,\AA) of star 5 would correspond to an M-type giant or supergiant if it were located in the cluster, while it is classified as a K1--2 type star,
%[SPECTRAL TYPE TO BE CONFIRMED] 
and it is likely a field star. This leaves no OB stars from which to determine the fundamental cluster parameters. Fitting the CCD and CMD yields $A_K=2.1$\,mag ($A_V=19.3$, with possibly significant differential reddening) and an upper limit of 7 Myr for the cluster age. \new{Our spectrophotometric distance of $\sim4$\,kpc corroborates the average kinematic distance of 4.9\,kpc to the star forming complexes near the cluster (Russeil \cite{Ru03}), considering the uncertainties.} VVV\,CL073's PDMF slope is compatible with the Salpeter slope. \new{The observed mass is $800\pm200$\,$M_\odot$ (using 52 MS stars), and we extrapolate a lower limit for the total mass of 1400\,$M_\odot$.}

One could suppose that this apparent group of stars may be the result of a low-extinction window. However, we consider this quite unlikely, since the two WR stars are located right at the postulated centre, where we expect to find the most massive stars in a cluster. Instead, we support the idea that it is either heavily reddened or has an unusual PDMF, with a few massive stars and many low-mass stars with much fainter magnitudes. One has to note that VVV\,CL073-2 could also be classified O4--6If. \new{When this spectral type is assumed, the distance to the star ($5.9\pm0.1$\,kpc) still corresponds to the cluster distance.}

\subsection{VVV\,CL074}

The background of this cluster is very complex, probably because of the presence of a dark cloud towards the North-East of the cluster, and the $S_{\rm B}$ level is quite uncertain. We therefore assume an average background value, while accounting for the strong variations as a function of radius. This allowed us to fit a King profile to the RDP, but with uncertain results. Nevertheless, we obtain a central position of R.A. (J2000)\,=\,16:32:05.86 and Dec (J2000)\,=\,$-47$:49:30.50, $S_{\rm B}=150\pm20$\,stars arcmin$^{-2}$, $S_0=880\pm114$\,stars arcmin$^{-2}$ and $r_{\rm{c}}=0.12'\pm0.02'$. The cluster radius is $0.6'\pm0.2'$ (i.e. $1.1\pm0.5$\,pc).

The MS observed in the CMD of this cluster appears quite broad. The CCD shows evidence of some (but not too severe) differential reddening. Interestingly, the reddening-free CMD also shows significant scatter in the $Q_{\rm NIR}$ distribution\new{, almost as if two sequences were present}. Among our spectroscopic targets, the OB stars are found at $-0.35 \leq Q_{\rm NIR} \leq -0.15$\,mag, while cool stars and star 9 (WN7 or O4--6If) attain higher values of $Q_{\rm NIR}$. One has to note that the classification of VVV\,CL074-9 is still ambiguous, since it could be WN7 or O4--6If. \new{The distance to that star would be $6.8\pm0.3$\,kpc if it is an O supergiant star, which is within the cluster considering the uncertainties.}
%[IS THIS COMING FROM WEIRD STARS OR   BAD DECONT OR SOMETHING ELSE...] 
Two cool stars are present in the CMD (stars 8 and 11) which could be cluster RSGs. While this is plausible for star 11, which has EW(CO)\,=\,19.2\,\AA\, and is classified as M1--2I, it is not for star 8, which has EW(CO)\,=\,14.7\,\AA\, and is classified as M3III. 
%[SPECTRAL TYPE TO BE   CONFIRMED] 
Consequently, only star 11 is considered a likely cluster member. 
%[SOMETHING ABOUT MIR]. 
\new{Using all OB stars, i.e. stars 1, 4, 5, 6 and 7, we get $A_K=2.78\pm0.01$\,mag ($A_V=23.4\pm0.1$\,mag) and a distance of $6\pm1$\,kpc. The luminosity class III has been attributed to the OB stars, since their spectra show that they are clearly not MS stars, and because their distance would have been very close to the edge of the Milky Way if they were supergiants. Moreover, this spectrophotometric distance is compatible with a kinematic distance of 4.9\,kpc (determined using the same regions as for VVV\,CL073).} This cluster does not seem to contain many PMS stars, if any. However, it is possible that PMS stars cannot be clearly isolated in the $Q_{\rm NIR}$ diagram because of the large uncertainties described above. The range of possible ages for this cluster is 4--5.5\,Myr. The PDMF slope of VVV\,CL074 is highly unreliable, due to the small mass-sampling region, hence the Salpeter slope ($\Gamma$=$-$1.35) was imposed. The observed mass is $1900\pm200$\,$M_\odot$ (using 170 MS stars), and the lower limit of the total cluster mass is 2500\,$M_\odot$.

\begin{figure}
\centering
\includegraphics[width=3cm,angle=90]{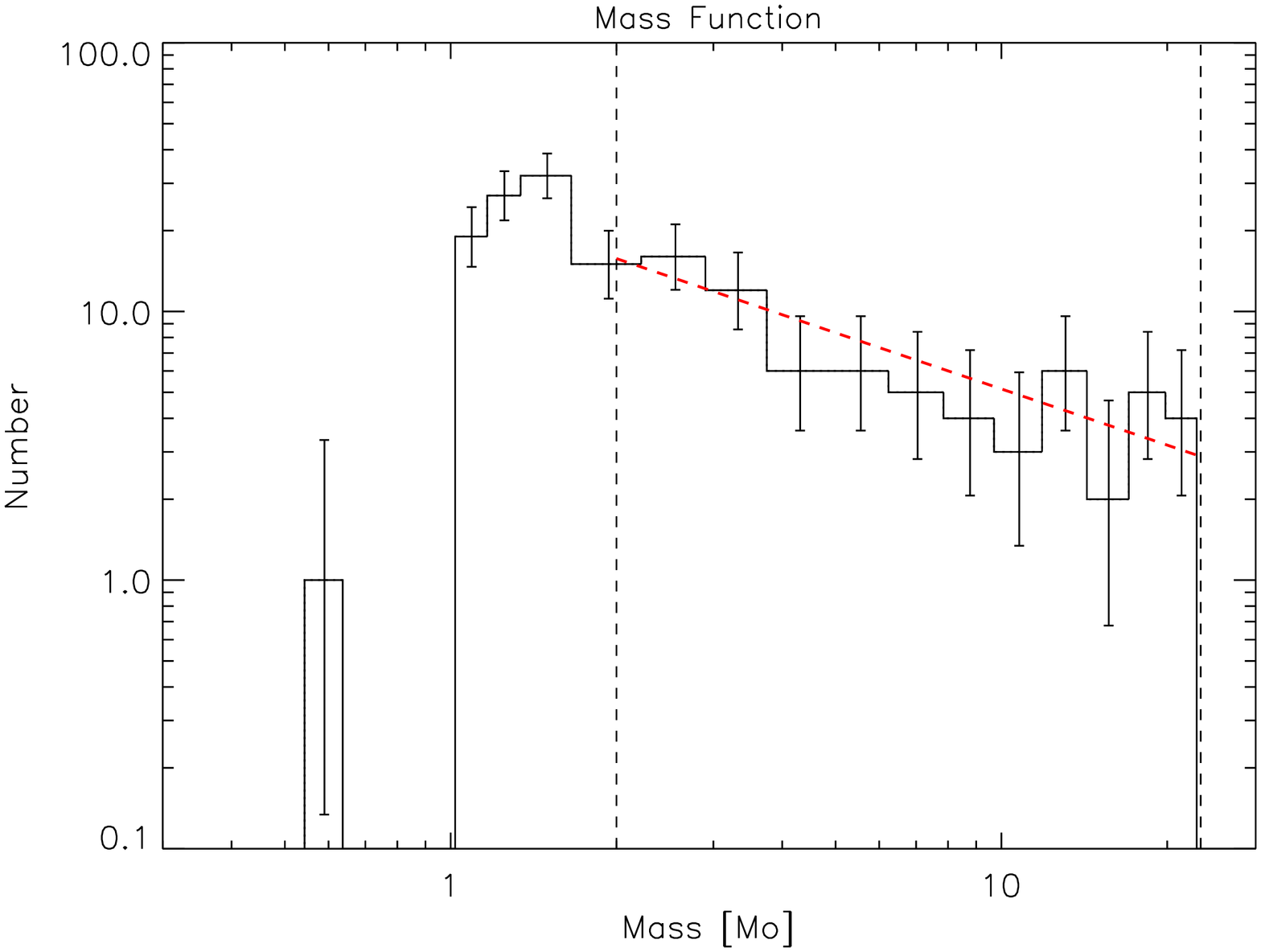}
\includegraphics[width=3cm,angle=90]{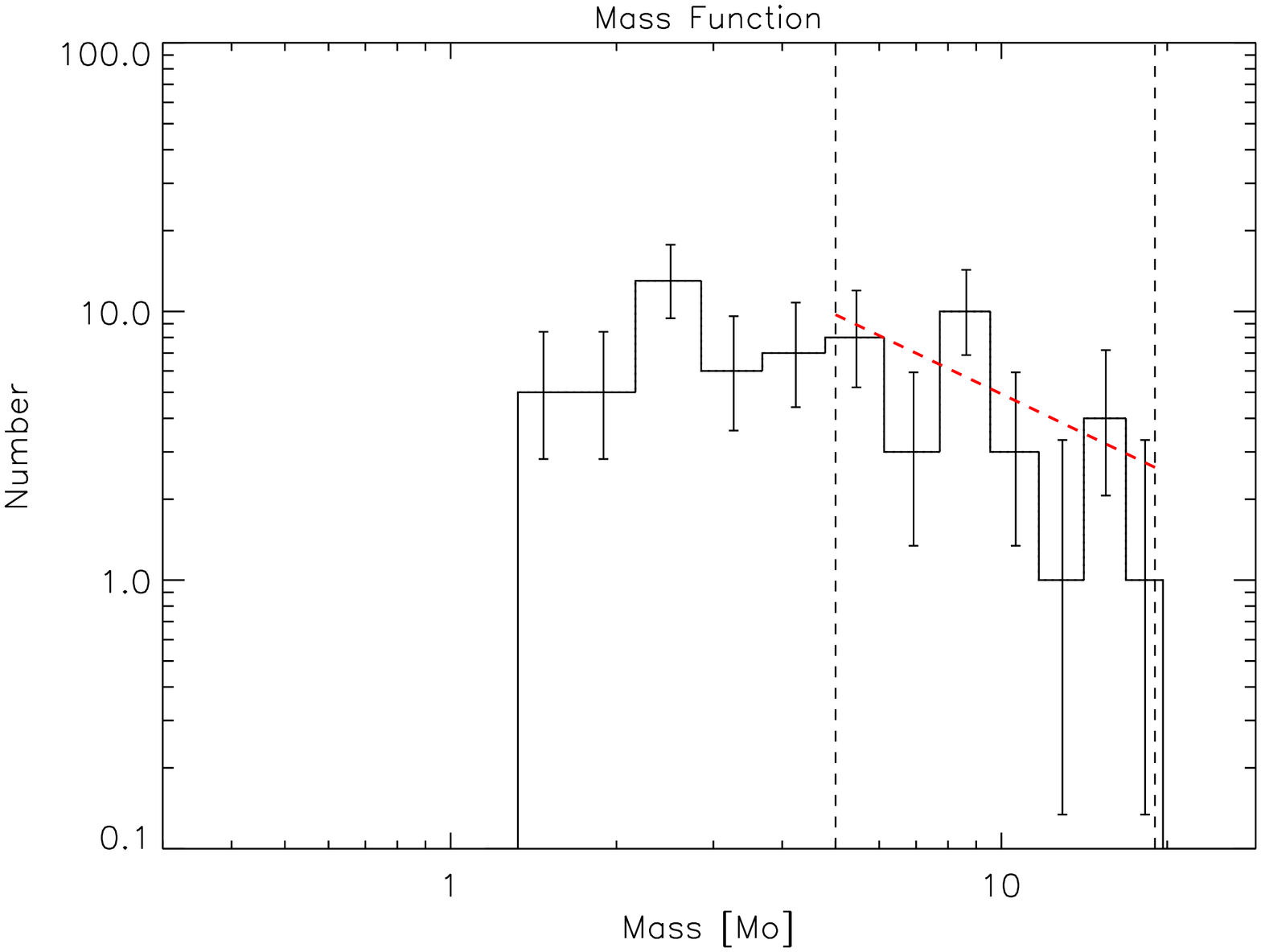}
\hspace{.5cm}
\includegraphics[width=3cm,angle=90]{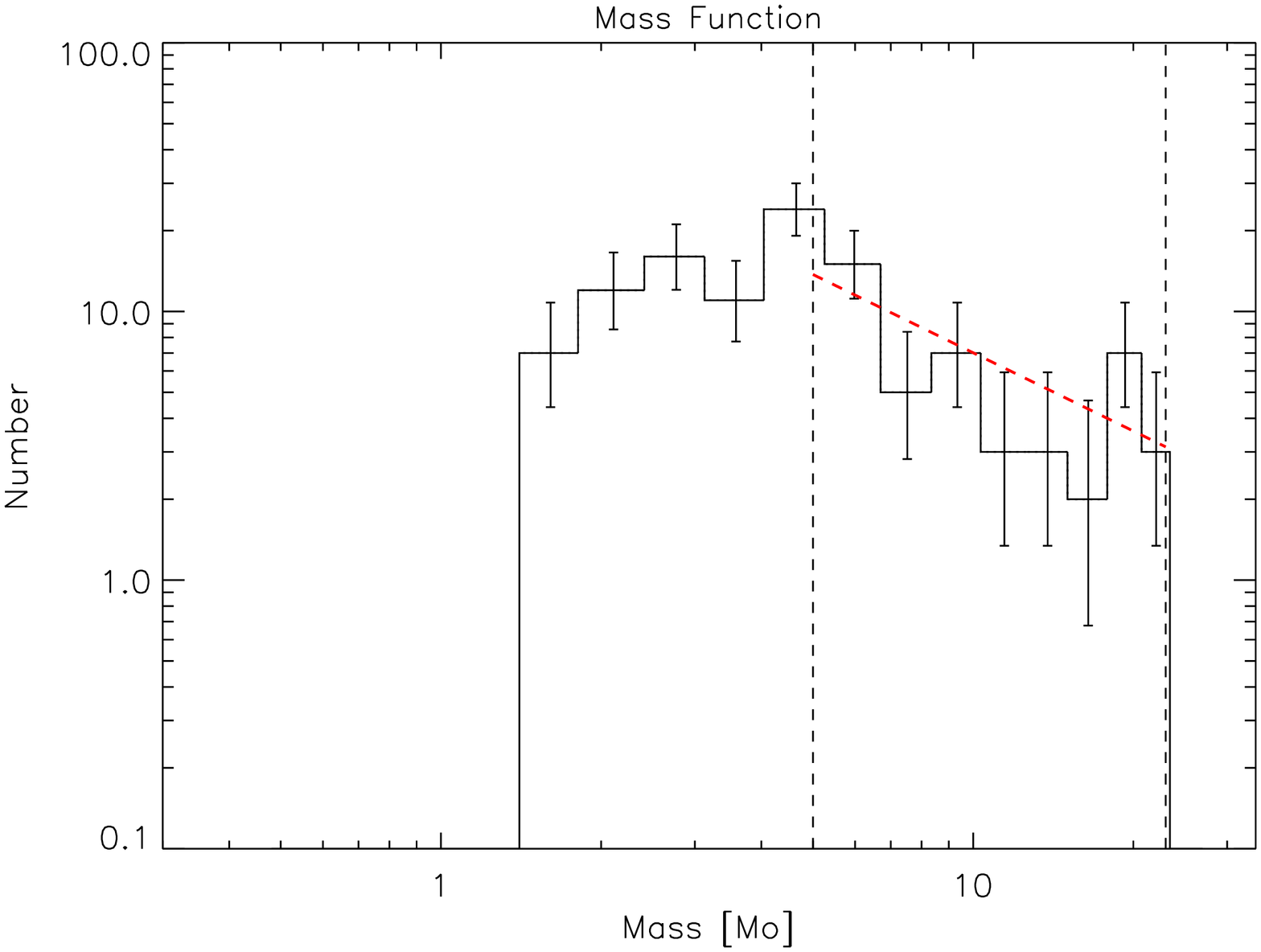}
\includegraphics[width=3cm,angle=90]{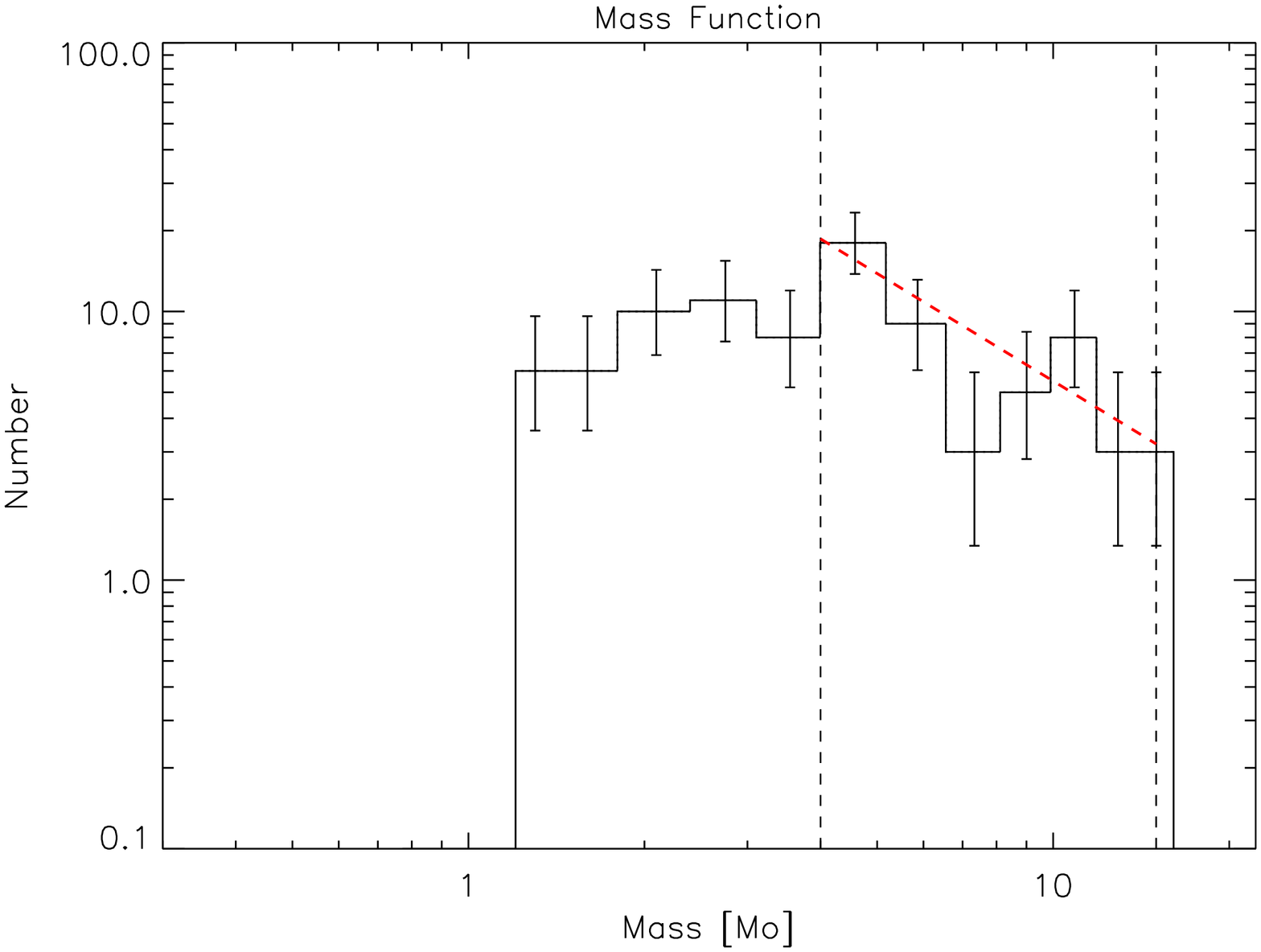}
\hspace{.5cm}
\includegraphics[width=3cm,angle=90]{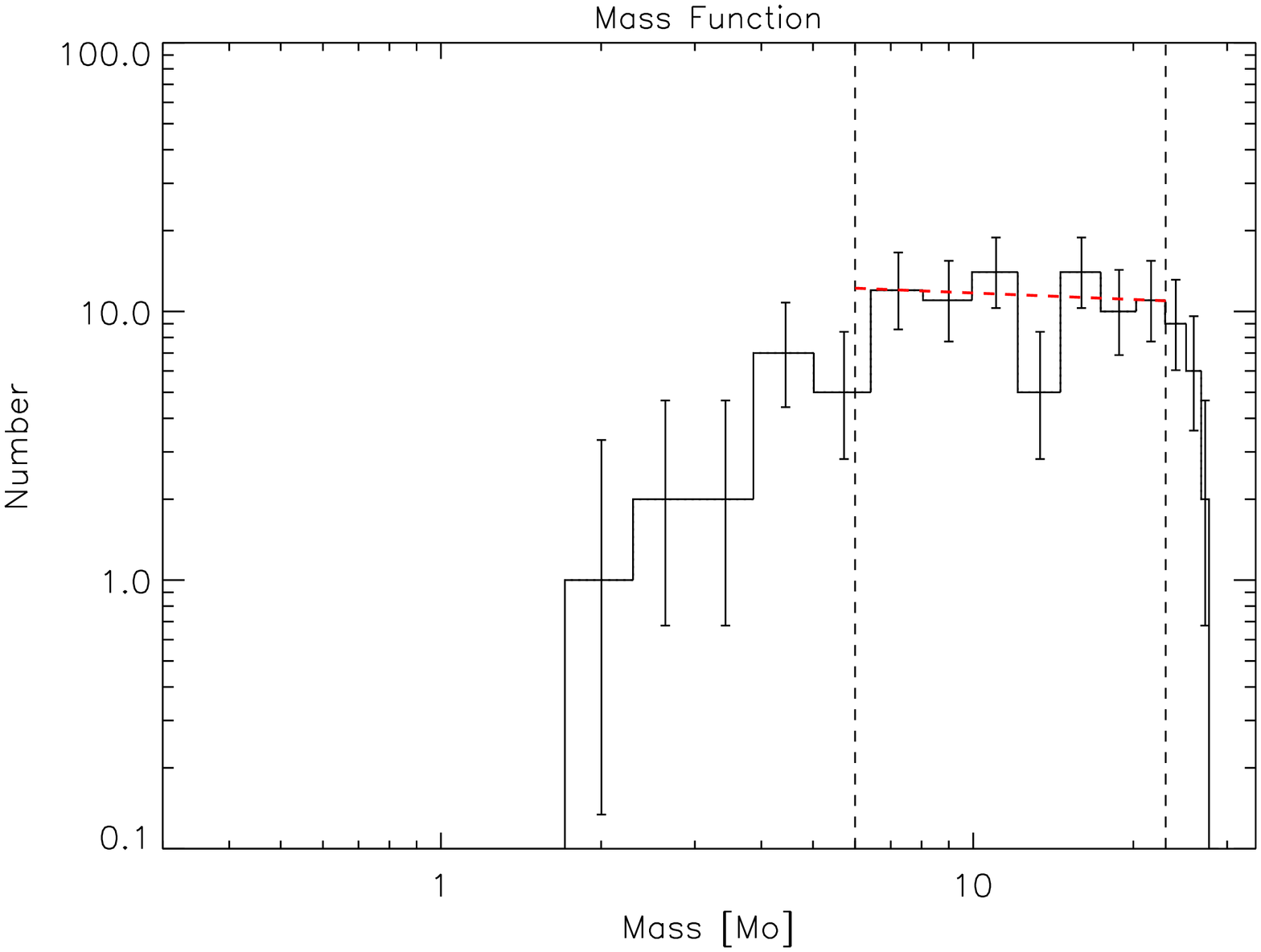}
\includegraphics[width=3cm,angle=90]{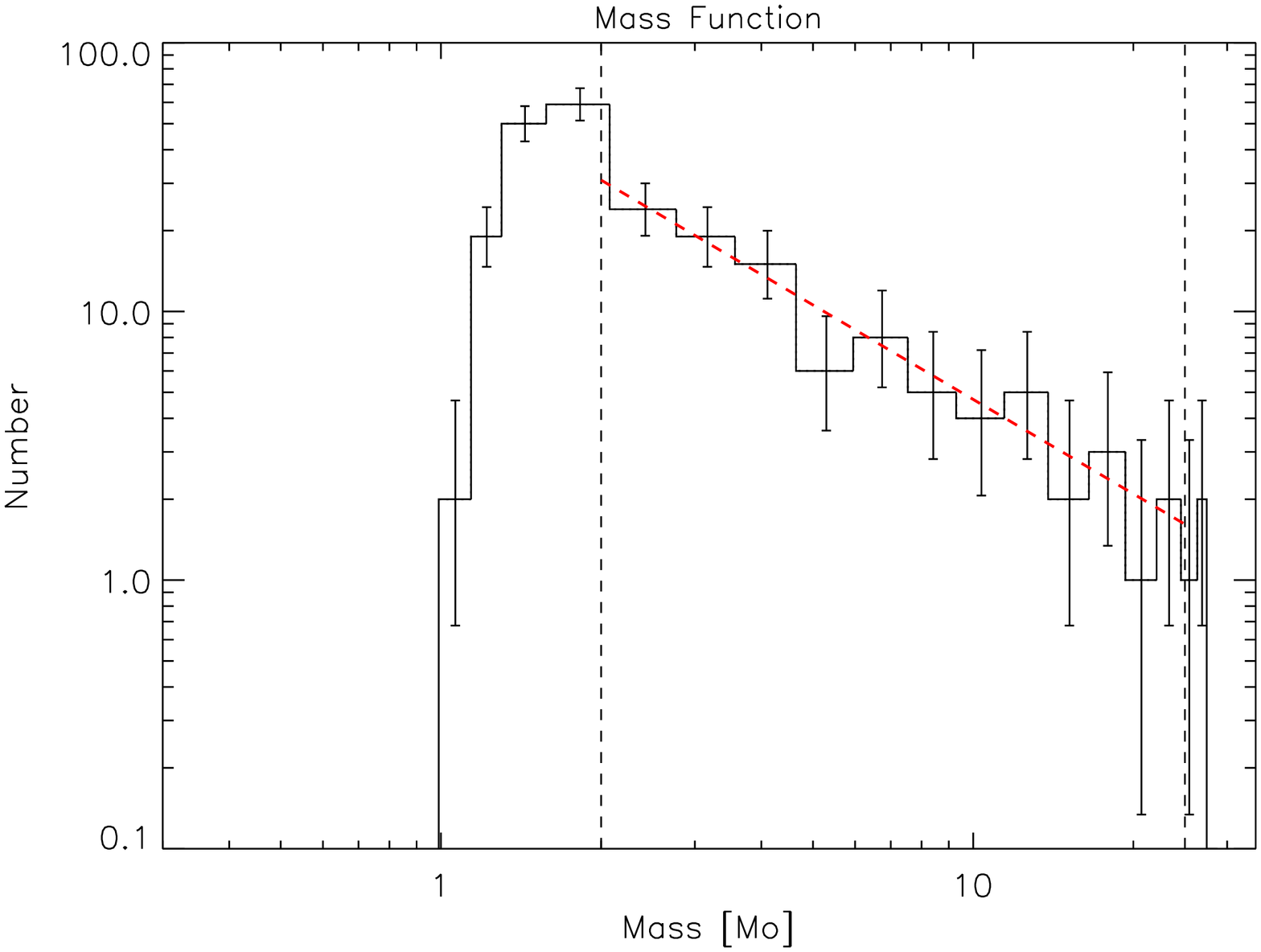}
\caption{PDMFs of the VVV CL009, VVV CL011, VVV CL036, VVV CL073, VVV CL074, VVV CL099. The slope is calculated between the mass limits denoted by vertical dashed lines. Overplotted as a red dashed line is the least-squares fitted slope.}
\label{FigMF1}%
\end{figure}

\begin{table}[!ht]\tiny
\caption{Summary of the physical cluster parameters.}
\label{results}
\begin{minipage}{180mm}
\begin{tabular}{ccccccr}
\hline
VVV    & radius & $A_V$ & d &   age     &  $\Gamma$&\multicolumn{1}{c}{Mass} \\
  CL     &   ($'$)    &   (mag) & (kpc)   & (Myr) & &\multicolumn{1}{c}{($M_\odot$)}  \\  
\hline
\vspace{.1cm}
009   & $0.6\pm 0.2$  &   $5.4\pm 0.1$  &  $5\pm 1$     &  4--6   &$-0.7\pm0.2$  &  660$\pm150$  \\  
\vspace{.1cm}
011   & $0.5\pm 0.1$  &   $9.4\pm 0.1$    &  $5\pm2$       &  3--7   &$-1.0\pm0.6$  &  390$\pm150$   \\ 
\vspace{.1cm}
036   & $>1.$         &   $24\pm 3$ &  $2\pm 1$     &  5--7   &$-1.0\pm0.4$  & 1800$\pm300$  \\  
\vspace{.1cm}
073   & $0.5\pm 0.5$  &  $\sim19.3$     &   $\sim4$         &  $<7$   &$-1.33\pm0.5$  & 800$\pm200$  \\  
\vspace{.1cm}
074   & $0.6\pm 0.2$  &  $23.4\pm 0.1$  &   $6\pm1$        &  4--6   & Salpeter  & 1900$\pm200$     \\ 
\vspace{.1cm}
099   & $0.8\pm 0.2$  &  $15.2\pm 0.2$  &   $4\pm 1$        &  4--6   &$-1.2\pm0.2$  & 1000$\pm100$   \\ 
\hline
\end{tabular}
\end{minipage}
\tablefoot{Columns include the VVV\,CL name, the cluster radius ($r_{\rm{c}}$), its reddening, distance, age, the slope of the PDMF (``Salpeter'' means that the slope has been imposed) and the observed mass.}
\end{table}

\subsection{VVV\,CL099}

The RDP of VVV\,CL099 is not like the RDP of a perfectly symmetric and nicely shaped cluster. Nevertheless, we could obtain a central position of R.A.  (J2000)\,=\,17:14:25.78, Dec (J2000)\,=\,$-38$:09:58.28. Filtering of the catalogue enabled us to obtain a better-defined profile, but our fit of a King profile was not fully satisfactory. We can still use the values we obtained as first estimates, with $S_{\rm B}=400\pm30$\,stars arcmin$^{-2}$, $S_0=638\pm67$\,stars arcmin$^{-2}$ and $r_{\rm{c}}=0.37'\pm0.08'$. The cluster radius is $0.8'\pm0.2'$ (i.e. $1.2\pm0.5$\,pc).

Stars 3, 9, 10 and 11 can be rejected based on their position in the CMD, combined with their spectral types. Moreover, the cool stars 9, 10 and 11 cannot be classified as giants or supergiants, since their EW(CO) values (15.8, 9.4 and 11.6\,\AA, respectively) would not correspond to their spectral types (M2--4V) in that case. Similarly, star 12 has EW(CO)\,=\,16.8\,\AA\, and is classified as M2--4V too.
%[SPECTRAL TYPES TO BE CONFIRMED] [SOMETHING ABOUT MIR] 
Other stars, excluding the WR stars stars 4, 5 and 7, can be used to derive the cluster parameters. We obtain $A_K=1.79\pm0.02$\,mag ($A_V=15.2\pm0.2$\,mag) and a distance of $4\pm1$\,kpc. Stars with $Q_{\rm NIR}=0.49$\,mag are defined as PMS stars. Interestingly, while our OB spectroscopic targets are close to the MS isochrone in the reddening-free CMD, the rest of the MS seems shifted towards higher $Q_{\rm NIR}$ values. However, the modelled MS can be aligned to the observed sequence if star\,8 is ignored for the determination of the cluster parameters. The fit of the PMS isochrones coincides with ages older than 5\,Myr, while the fit of MS isochrones yields an age between 4 and 6\,Myr, which are mutually consistent. The PDMF slope is compatible to the Salpater slope and \new{the observed mass, determined using 168 MS stars, is $1000\pm100$\,$M_\odot$. This gives a lower limit for the total extrapolated mass of 1600\,$M_\odot$.}

\section{Discussion}\label{Discussion}

\subsection{On the likelihood of these objects being real clusters}\label{realclust}

The new stellar groups discovered by Borissova et al. (\cite{Bo11}) exhibit significant spatial stellar overdensities. However, only a follow-up study can confirm if these are real clusters or simple asterisms. This study covers six of these groups in which spectroscopic observations have revealed the presence of at least one WR star, sometimes associated with as many as 5--9 OB stars. The presence of this many OB and WR stars inside such a small angular radius is an indication of clustering. It cannot be completely excluded that O stars are formed outside a cluster environment, since a certain fraction ($\sim$\,4\%) are observed to be apparently isolated (de Wit et al. \cite{de05}). One could then wonder if a random distribution of field O stars can produce an alignment which may give the impression of clustering. However, it seems that this same fraction of O stars are formed in clusters which contain no other massive stars, giving the impression of being isolated, when stars are sampled randomly from a standard PDMF and a standard cluster mass function (Parker \& Goodwin \cite{Pa07}). Finally, field O stars ejected from nearby clusters (runaways) are the only real isolated stars and should always exhibit peculiarly high proper motions and/or RVs (e.g. Gvaramadze et al. \cite{Gv12}).

In four of our sample clusters, 5--9 OB stars have spectrophotometric distances of up to 1\,kpc, i.e. within the uncertainties of the distances determined for their associated clusters (note that these uncertainties are dominated by the uncertainties in the spectral classification). Moreover, all OB stars in a given cluster are found along the same MS in the CMDs and are never separated by an angular distance greater than $\sim1'$. Considering the distances to the clusters, this means that all OB stars are confined within 1--2\,pc. As for the other two clusters (VVV\,CL011 and VVV\,CL073), they contain one B and one WR star, and two WR stars, respectively, and are confined within less than 1\,pc. All six clusters have WR stars near their centers, i.e. where massive stars are preferentially found when clusters are compact (Bate \cite{Ba02}; Allison et al. \cite{Al09}). Can a random distribution of massive field stars mimic what seem to be typical cases of massive, compact clusters?

Let us, for a moment, assume that O stars can form in isolation, and have a look at how O- and B-type stars would be distributed in the Galactic plane when one randomly picks stars from a standard PDMF along the lines of sight to our sample clusters. Using the Besan\c{c}on model of Galactic stellar population synthesis (Robin et al. \cite{Ro03}), we looked at the distribution of stars with spectral types between O7 and B5 (more or less the spectral types observed in this study) with distances between 0 and 50\,kpc. Because we are studying clusters at low Galactic latitudes, we could in principle use highly diffuse extinction (i.e.\,$\>0.7$\,mag kpc$^{-1}$) for our model, but in case we are dealing with windows of lower extinction, we used normal to low-level diffuse extinction and look for the upper limit of the number of OB stars per unit surface area. The simulation was run using 35 patches of sky, covering 11\,arcmin$^2$ each, around each of the clusters' coordinates. On average, we obtained a total of $0.2\pm 0.1$\,O- and B-type stars arcmin$^{-2}$, which is well below what is needed to reach the densities observed (i.e. values of 2.5--8 O and B stars arcmin$^{-2}$, depending on the cluster). We hence affirm that the six stellar groups studied here are not asterisms, but real young, massive and compact open clusters.

\subsection{Milky Way structure}

\begin{figure}
  \centering
   \includegraphics[width=9.0cm]{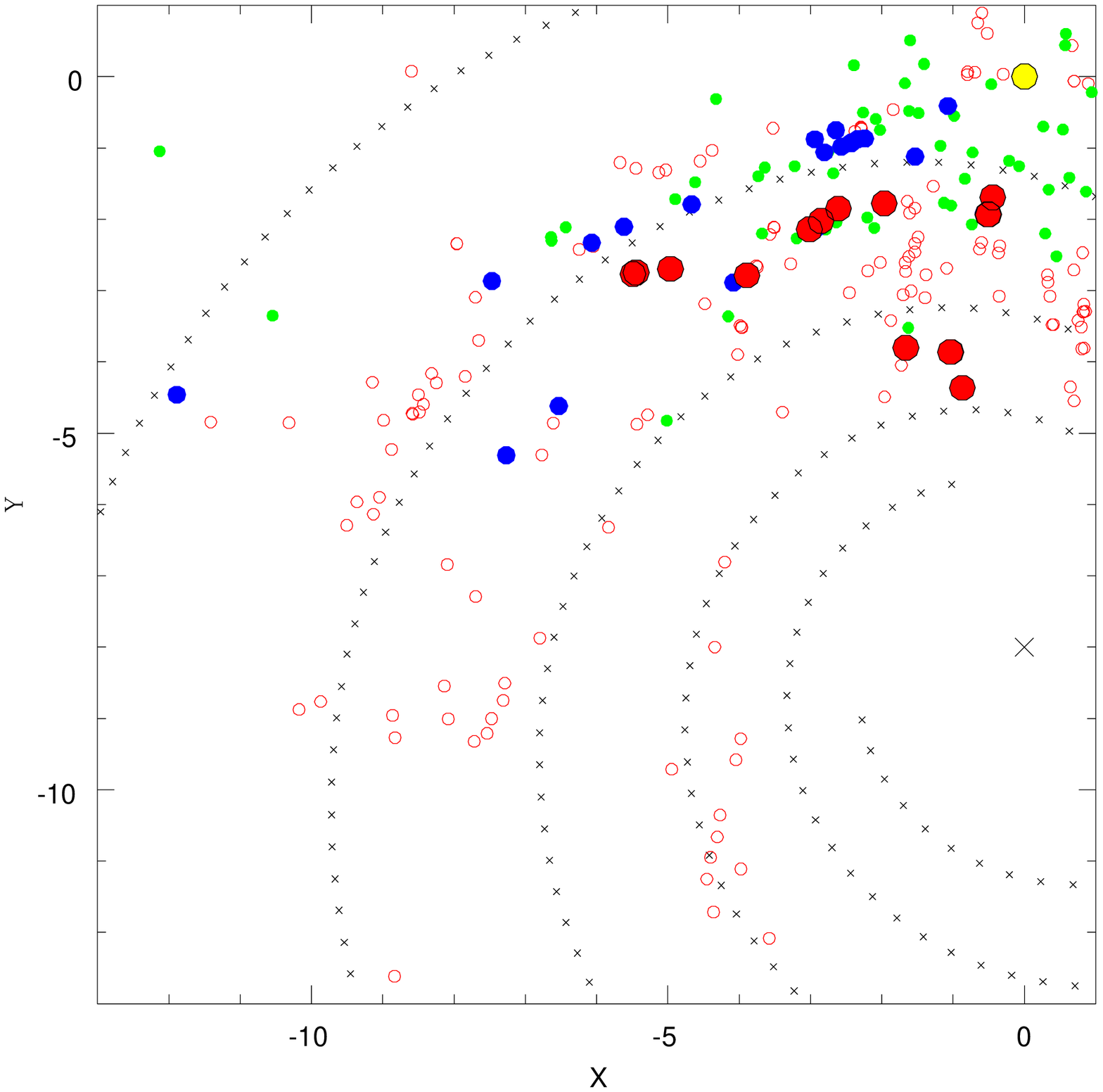}
   \caption{Comparison of the observations to the Vall\'ee (\cite{Va08}) model with four arms in the fourth quadrant of the Galaxy. We assumed a distance to the Galacic Centre of $R_0$ = 8.0 kpc. Cepheid distances (Majaess et al. \cite{Ma09}) are plotted as green points, open clusters and young groups from optical ($UBVI$) data (from several studies) are plotted as blue points, while red points show near-IR clusters studied using VVV data (Paper I, this study and Baume et al. \cite{Ba10}). Red circles mark the H{\sc ii} tracers (Hou et al. \cite{Ho09}) and the yellow dot marks the position of the Sun.}
              \label{spiral}%
\end{figure}

In this section we discuss the spiral structure in the fourth quadrant of the Milky Way based on the location of objects related to very young populations ($<10$\,Myr) which can be considered reliable tracers of spiral structure. In Figure\,\ref{spiral} we show a hybrid map of this region of the Galaxy, plotting our results in combination with data from previous papers. We have included the embedded clusters studied in this paper, as well as those from Paper I and preliminary results from Baume et al. (\cite{Ba10}). We also include optical results from studies of young open clusters and young populations near the Carina region (V\'azquez et al. \cite{Va05}; Baume et al. \cite{Ba09}; Carraro \& Costa \cite{Ca09}) and long-period Cepheids (see Majaess et al. \cite{Ma09}). We also plot the H{\sc   ii}-region compilation of Hou et al. (\cite{Ho09}) and the four-arm Galactic model of Vall\'ee (\cite{Va08}). We adopt 8\,kpc as the distance to the Galactic Centre.

\begin{figure}
  \centering
   \includegraphics[width=9.0cm]{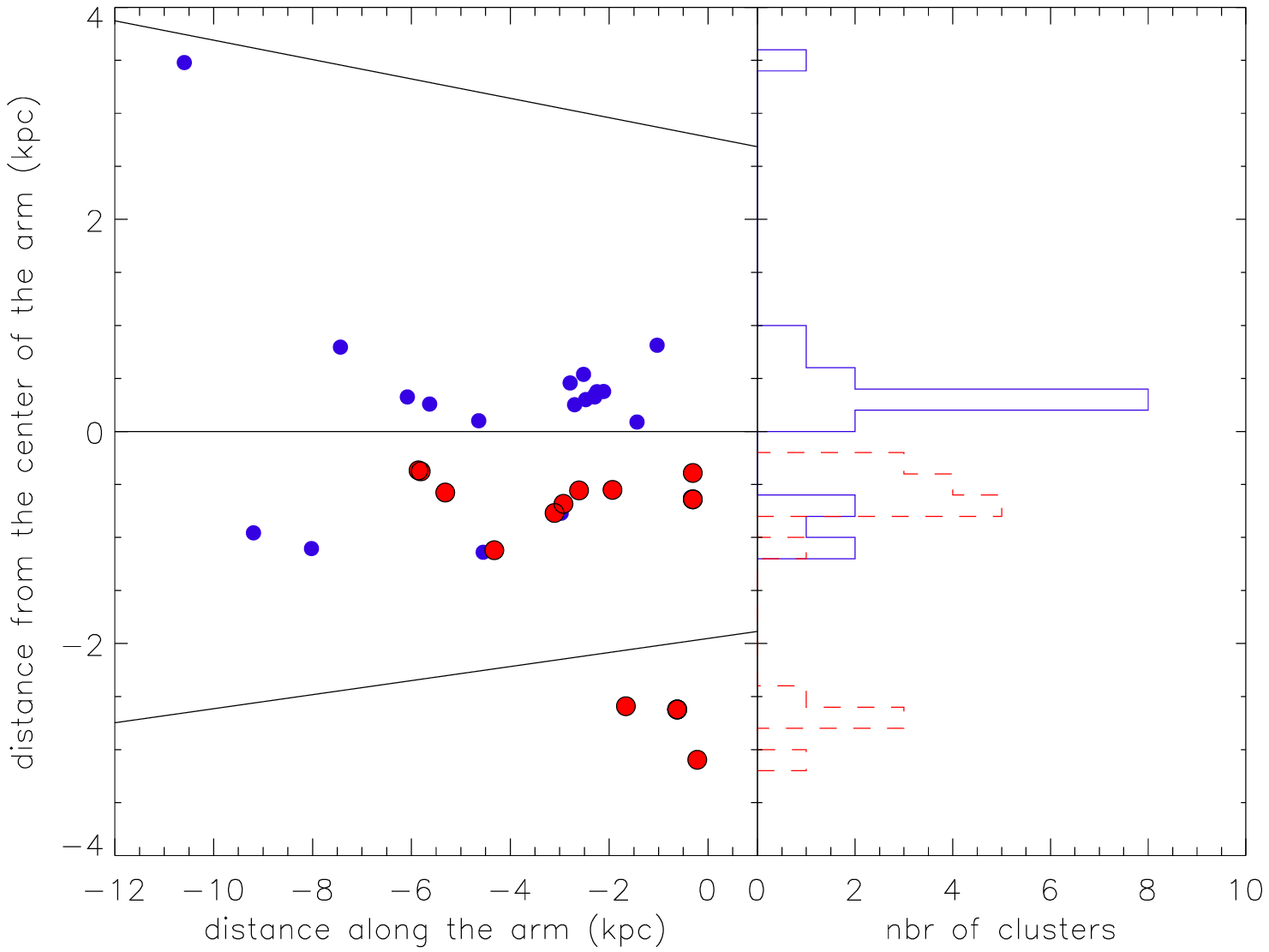}
   \caption{{\it Left}:~Distance from the center of the Carina arm (as modeled by VallŽe \cite{Va08}) of the clusters shown in Figure\,\ref{spiral} as a function of their position along the arm. The Centaurus (over) and the Crux (under) arms are also plotted as slopes. {\it Right}:~Distribution of the clusters with respect to the center of the Carina arm. The distribution of the clusters from optical data is plotted in blue, and the distribution of the clusters studied using VVV is plotted in red.}
              \label{distcl}%
\end{figure}

Considering only the clusters and young populations, three clear groups can be noticed: a trace defined mainly by clusters detected in the optical regime, another parallel trace of clusters studied in the IR based on VVV data and finally a group of embedded clusters closer to the Galactic Centre. The first two traces follow the spiral structure indicated by H{\sc ii} regions and Cepheids, but they seem offset with respect to each other. In addition, none match the Vall\'ee model, which places the Carina arm between them. This offset is real and cannot have been caused by systematic effects, since a systematic error on the distance to the clusters (the only source of uncertainties in this analysis) would not preserve the spiral shape. There are two interpretations for the difference between the traces of the clusters detected in the optical and those discovered using the VVV data. First, both may trace two different Galactic arms (Carina and Centaurus, as indicated in Paper I) and the Vall\'ee model would consequently need different parameters (pitch angle and/or interarm distance) to fit the fourth quadrant. Second, if we assume that the Vall\'ee model is right, both traces may reveal the internal structure of only one arm (the Carina arm), as can be seen in other galaxies (e.g. M51). Clusters discovered in the VVV data would be embedded in dark clouds behind the young (optical) populations at the front end of the arm. \new{Figure\,\ref{distcl} illustrates the projected cluster distribution along the Carina arm, reinforcing the second interpretation. However, there is evidence of discrepancies among cluster distances obtained using different methods and this could affect the interpretation of the data (see Moises et al. \cite{MD11} for a full discussion).}

\subsection{The 11 newly discovered WR stars}\label{newWR}

Here we provide a summary of the 11 stars showing WR-like emission lines that were discovered in our sample clusters. These stars are VVV\,CL009-6 (WN7/Of), VVV\,CL011-2 (OIf or WN9), VVV\,CL036-9 (WN6), VVV\,CL073-2 and -4 (O4--6If or WN9 and WN7, respectively), VVV\,CL074-2, -3 and -9 (WC8, WN8 and O4--6If or WN7, respectively), and VVV\,CL099-4, -5 and -7 (WN6, WN6 and WC8, respectively). The three stars which were classified with an ambiguous spectral type due to the similarities of the spectral lines of Of and WR stars in the $K$-band will be studied further in an upcoming paper (A.-N. Chen\'e et al., in prep), and will be consider as WR for the rest of this sections. \new{We remark that if all these stars are O supergiants, VVV\,CL073-2 and VVV\,CL074-9 would be both close to and within their host cluster, respectively.} Following the nomenclature used in the VIII$^{\rm{th}}$ catalogue of WR stars\footnote{Crowther, Baker \& Kus   2012,\\ http://pacrowther.staff.shef.ac.uk/WRcat/history.php}, we can attribute a WR number to them. Table\,\ref{WR} includes some additional information that is not already included in Table\,\ref{TabSp}.

\begin{table}[!ht]
\caption{Additional information about the new WR stars}
\label{WR}
\begin{minipage}{180mm}
\begin{tabular}{ccccrc}
\hline
 & name    &  WR  &  $l$    &   \multicolumn{1}{c}{$b$}  &   d     \\
 &   VVV    & num. & (deg) & \multicolumn{1}{c}{(deg)} & (kpc) \\
\hline
\noalign{\smallskip}
\hline
\noalign{\smallskip}
\multicolumn{4}{l}{\it VVV\,CL009:} &&\\
 & 6  & 45d & 296.7623 & --1.1086 & $5\pm 1$\\
\multicolumn{6}{l}{}\\
\multicolumn{4}{l}{\it VVV\,CL011:} &&\\
 & 2  & 46ae & 298.5065 & --0.1702 & $5\pm2$\\
\multicolumn{6}{l}{}\\
\multicolumn{4}{l}{\it VVV\,CL036:} &&\\
 & 9  & 60aa & 312.1248 & 0.2135 & $2\pm 1$\\
\multicolumn{6}{l}{}\\
\multicolumn{4}{l}{\it VVV\,CL073:} &&\\
 & 2  & 75bda & 335.8921 & 0.1327 & $\sim4$ \\
 & 4  & 75bdb & 335.8925 & 0.1321 & $''$ \\
\multicolumn{6}{l}{}\\
\multicolumn{4}{l}{\it VVV\,CL074:} &&\\
 & 2  & 75bfa & 336.3755 & 0.1987 & $6\pm 1$ \\
 & 3  & 75bfb & 336.3731 & 0.1957 & $''$ \\
 & 9  & 75bfc & 336.3734 & 0.1942 & $''$ \\
\multicolumn{6}{l}{}\\
\multicolumn{4}{l}{\it VVV\,CL099:} &&\\
 & 4  & 84f & 348.7265 & 0.3276 & $4\pm1$ \\
 & 5  & 84g & 348.7277 & 0.3256  & $''$ \\
 & 7  & 84h & 348.7274 & 0.3244 & $''$ \\
\hline
\end{tabular}
\end{minipage}
\end{table}

The 11 WR stars have late spectral type, which corresponds to the Galactic population (cf Crowther \cite{Cr07}). Also, none of them have a WNLh spectral type (like in Danks\,2, see Davies et al. \cite{Da12} and Paper\,I), imposing a minimal age of 2--3\,Myr for our six sampled clusters. According to the population of O stars found in the clusters, we can deduce that the precursors of the WR stars presented here were intermediate MS O stars (i.e. typically O5--8V), corresponding to initial masses $\sim$30--50\,$M_\odot$. Of course, more details will be obtained after the spectral analysis are presented in the next study.
%Peut-etre discuter la completude? Si les echantillons sont complets, les rapports WC/WN dans chaque amas contraignent la metallisite et/ou le mode de formation (burst ou formation continu). CF Schearer & Vacca (1998) ou Martins et al. (2007) par ex.

\section{Summary}\label{Summary}

This paper presents the first study of six new, young, massive clusters from the VVV catalog (Borissova et al. \cite{Bo11}). These clusters contain at least one WR (the VVV\,CL011-2 could be O4--6If star) star, and one or more O- and B- type stars. They are highly reddened (A$_V\sim 5-24$\,mag) and compact ($\sim 1-2$\, pc). Their age is less than 7\,Myr and they have total masses between 0.8 and 2.2 $10^{3}$\,$M_\odot$. It is highly improbable that the six stellar groups are not real clusters but instead a simple alignment of O- and B- type stars along the line-of-sight.

We report the discovery of a total of 8 WR stars, and 3 stars showing WR-like emission line which could be either WNL or OIf stars. From a preliminary study of the clusters' population of WR and O stars, we estimate the initial mass of the most massive star in the sampled clusters was 30--50\,$M_\odot$ and determine an minimum age limit of 2--3\,Myr for all six clusters. The clusters VVV\,CL036 and VVV\,CL074 contain one or more WR star and probably a least one M supergiant star. This cohabitation will be discussed in an upcoming publication. 

Finally, the six sampled clusters, joined with the previously studied VVV clusters, were used to trace the spiral structure of the Milky Way. Interestingly, the VVV clusters follow the spiral structure indicated by H{\sc ii} regions and Cepheid stars, but are clearly offset with respect to the open clusters discovered in optical light. The two possible explanations are that both optical and VVV clusters trace different Galactic arms, or that they trace different parts of the same arm, in this case, the Carina arm.
 
\begin{acknowledgements}
We would like to thank the anonymous referee for his help to improve this work. This project is supported by the Chilean Ministry for the Economy, Development, and Tourism's Programa Iniciativa Cient\'{\i}fica Milenio through grant P07-021-F, awarded to The Milky Way Millennium Nucleus. ANC, DG and DM gratefully acknowledge support from the Chilean Centro de Astrof\'{\i}sica FONDAP No. 15010003 and the BASAL Chilean Centro de Excelencia en Astrof\'{\i}sica y Tecnolog\'{\i}as Afines (CATA) PFB-06. ANC also acknowledges the Comite Mixto ESO-Gobierno de Chile. JB is supported by FONDECYT No. 1120601 and RK by the Centro de Astrof\'{\i}sica de Valpara\'{\i}so and Proyecto DIUV23/2009. Support for MC is also provided by Proyecto Fondecyt Regular \#1110326, and Anillo ACT-86. RdG acknowledges partial research support through grant 11073001 from the National Natural Science Foundation of China. MSNK is supported by a Ci\^encia 2007 contract, funded by FCT/MCTES (Portugal) and POPH/FSE (EC). The data used in this paper have been obtained with NTT/SofI at ESO's La Silla Observatory, and with Clay/MMIRS at Las Campanas Observatory. This research has made use of the SIMBAD database, operated at CDS, Strasbourg, France. We gratefully acknowledge use of data from the ESO Public Survey programme ID 179.B-2002 taken with the VISTA telescope, and data products from the Cambridge Astronomical Survey Unit.
\end{acknowledgements}

\end{document}